\documentclass[pre,twocolumn]{revtex4-2}
\usepackage{amsmath}
\usepackage{amssymb}
\usepackage{graphicx}
\usepackage{braket}
\usepackage{stmaryrd}
\usepackage{hyperref}
\usepackage[usenames,dvipsnames]{xcolor}
\usepackage{dsfont}

\hypersetup{
    colorlinks,
    citecolor=blue,
    linkcolor=blue,
    urlcolor=blue
}

\newcommand{\dblket}[1]{|#1\rangle \! \rangle}
\newcommand{\dblbra}[1]{\langle \! \langle #1|}
\newcommand{\dblbraket}[1]{\langle \! \langle #1 \rangle \! \rangle}

\begin{document}
\title{Exact Solution to Quantum Dynamical Activity}

\author{Tomohiro Nishiyama}
\email{htam0ybboh@gmail.com}
\affiliation{Independent Researcher, Tokyo 206-0003, Japan}

\author{Yoshihiko Hasegawa}
\email{hasegawa@biom.t.u-tokyo.ac.jp}
\affiliation{Department of Information and Communication Engineering, Graduate
School of Information Science and Technology, The University of Tokyo,
Tokyo 113-8656, Japan}
\date{\today}

\begin{abstract}
The quantum dynamical activity constitutes a thermodynamic cost in trade-off relations such as the quantum speed limit and the quantum thermodynamic uncertainty relation. However, calculating the quantum dynamical activity has been a challenge.
In this paper, we present the exact solution for the quantum dynamical activity by deploying the continuous matrix product state method. Moreover, using the derived exact solution, we determine the upper bound of the dynamical activity, which comprises the standard deviation of the system Hamiltonian and jump operators. 
We confirm the exact solution and the upper bound by performing numerical simulations.
\end{abstract}

\maketitle

\section{Introduction}

Uncertainty relations are pivotal relations that outline what is feasible or impractical in the real world. The most prominent example is Heisenberg's uncertainty principle, which shows the uncertainty associated with conjugate observables such as position and momentum operators \cite{Heisenberg:1927:UR,Robertson:1929:UncRel}. 
Another well-studied uncertainty relation in quantum systems is the quantum speed limit (QSL) \cite{Mandelstam:1945:QSL,Margolus:1998:QSL,Deffner:2017:QSLReview}.
The QSL corresponds to the energy-time uncertainty, which imposes a restriction on how quickly a quantum system can transition between states.
The concept of a speed limit was recently generalized in classical stochastic systems,
referred to as the classical speed limit (CSL) \cite{Shiraishi:2018:SpeedLimit,Ito:2018:InfoGeo,Ito:2020:TimeTURPRX,Nicholson:2020:TIUncRel,Vo:2020:TURCSLPRE,Vu:2022:OptimalTransportPRX}.
Recently, a closely related uncertainty principle known as the thermodynamic uncertainty relation (TUR) has been actively studied in the fields of stochastic and quantum thermodynamics \cite{Barato:2015:UncRel,Gingrich:2016:TUP,Garrahan:2017:TUR,Dechant:2018:TUR,Terlizzi:2019:KUR,Hasegawa:2019:FTUR,Erker:2017:QClockTUR,Carollo:2019:QuantumLDP,Hasegawa:2020:QTURPRL}. 
The TUR states that achieving higher accuracy should be accompanied by a higher thermodynamic cost. 
In CSLs and classical TURs, along with entropy production, the most studied thermodynamic cost is the dynamical activity \cite{Maes:2020:FrenesyPR}, which quantifies the activity of a Markov process (see Eq.~\eqref{eq:classical_DA_def}).
The dynamical activity plays a central role in the thermodynamic costs of CSLs \cite{Shiraishi:2018:SpeedLimit,Vo:2020:TURCSLPRE} and classical TURs \cite{Garrahan:2017:TUR,Terlizzi:2019:KUR}. 

The concept of classical dynamical activity is generalized to incorporate quantum effects and referred to as \textit{quantum dynamical activity}.
The quantum dynamical activity
is defined in the dynamics determined by the Lindblad equation [Eq.~\eqref{eq:GKSL_eq_def}] and
has recently been studied in quantum stochastic thermodynamics \cite{Hasegawa:2020:QTURPRL,Hasegawa:2023:BulkBoundaryBoundNC,Kewming:2023:FPT}. 
Suppose that the dynamics begin at $t = 0$ and end at $t = \tau >0$.
Let $\mathcal{B}(\tau)$ be the quantum dynamical activity within the interval $[0,\tau]$ (defined in Eq.~\eqref{eq:QDA_def}).
Consider the number of jump events in a continuous measurement.
The relative variance follows the quantum TUR \cite{Hasegawa:2020:QTURPRL}:
\begin{align}
    \frac{\mathrm{Var}[N(\tau)]}{\braket{N(\tau)}^{2}}\ge\frac{1}{\mathcal{B}(\tau)},
    \label{eq:QTUR1}
\end{align}
where $N(\tau)$ is a counting observable that counts the number of jump events within the interval $[0,\tau]$ and $\braket{N(\tau)}$ and $\mathrm{Var}[N(\tau)]$ are the mean and variance of $N(\tau)$, respectively. 
The quantum TUR of Eq.~\eqref{eq:QTUR1} states that a higher precision can be achieved provided that the system allows for more quantum dynamical activity. 
Moreover, $\mathcal{B}(\tau)$ appears in the QSL in a continuous-measurement setting.
Let $\rho(0)$ and $\rho(\tau)$ be the initial and final density operators governed by the Lindblad equation, and
let $\mathcal{L}_D\left(\rho(0), \rho(\tau)\right) \equiv \arccos \left[\sqrt{\operatorname{Fid}\left(\rho(0), \rho(\tau)\right)}\right]$ be the Bures angle between the initial and the final states,
where $\mathrm{Fid}\left(\rho_{1},\rho_{2}\right)\equiv\left(\mathrm{Tr}\sqrt{\sqrt{\rho_{1}}\rho_{2}\sqrt{\rho_{1}}}\right)^{2}$ is the quantum fidelity. 
Subsequently, the following QSL holds \cite{Hasegawa:2023:BulkBoundaryBoundNC}:
\begin{align}
    \mathcal{L}_{D}(\rho(0),\rho(\tau))\le\frac{1}{2}\int_{0}^{\tau}dt\,\frac{\sqrt{\mathcal{B}(t)}}{t}.
    \label{eq:QSL_B1}
\end{align}
As the Bures angle determines the distance between the two density operators, Eq.~\eqref{eq:QSL_B1} states that, for the system to change its state more, the system demands more quantum dynamical activity. 
Considering a closed quantum limit, Eq.~\eqref{eq:QSL_B1} is reduced to the Mandelstam-Tamm bound \cite{Mandelstam:1945:QSL}. 
The appearance of $\mathcal{B}(t)$ in the two distinct uncertainty relations, given by Eqs.~\eqref{eq:QTUR1} and \eqref{eq:QSL_B1}, is not a coincidence. 
These are the two aspects of the same geometric inequality \cite{Hasegawa:2023:BulkBoundaryBoundNC}.

Equations~\eqref{eq:QTUR1} and \eqref{eq:QSL_B1} show that the quantum dynamical activity $\mathcal{B}(t)$ serves as a fundamental cost in systems described by the Lindblad equation.
However, deriving its exact expression remains challenging because
the quantum dynamical activity was defined using the quantum Fisher information [Eq.~\eqref{eq:QDA_def}], which lacks a closed-form representation. 
In this study, we derive the exact solution for the quantum dynamical activity [Eq.~\eqref{eq:main_result1}].
The calculation is based on a continuous matrix product state (cMPS), which encodes the dynamics into the quantum field.
Moreover, we derive the upper bounds for the quantum dynamical activity, comprising the moments of the Hamiltonian and jump operators. 
We find that the upper bound is tight for short durations.
We validate the exact solution and the upper bound by performing numerical simulations. 

\section{Methods}
The quantum dynamical activity is defined in the Lindblad equation.
Let $\rho_S(t)$ be the density operator of the system of interest. 
The Lindblad equation is represented by \cite{Gorini:1976:GKSEquation,Lindblad:1976:Generators} $\dot{\rho}_S = \mathcal{L}\rho_S$, where $\mathcal{L}$ is the Lindblad superoperator defined by
\begin{align}
    \mathcal{L}\rho_{S}=-i[H_{S},\rho_{S}]+\sum_{m=1}^{N_{C}}\mathcal{D}\left[L_{m}\right]\rho_{S}.
    \label{eq:GKSL_eq_def}
\end{align}
Here, $\mathcal{D}[L] \rho_S=L \rho_S L^{\dagger}-\frac{1}{2}\left\{L^{\dagger} L, \rho_S\right\}$ denotes the dissipator. 
The Lindblad equation incorporates both the classical Markov process and the closed-quantum evolution.
The classical Markov process and closed quantum dynamics are represented by $H_S = 0$ and $L_m=0$, respectively,
in Eq.~\eqref{eq:GKSL_eq_def}. 
The Lindblad equation can be expressed using the Kraus representation:
\begin{align}
    \rho_{S}(t+dt)=\sum_{m}V_{m}(dt)\rho_{S}(t)V_{m}(dt)^{\dagger},
    \label{eq:Kraus_UM_def}
\end{align}
where $V_m(dt)$ denotes the Kraus operator:
\begin{align}
    V_{0}(dt)&\equiv\mathbb{I}_{S}-idtH_{\mathrm{eff}},\label{eq:Kraus_V0_def}\\
    V_{m}(dt)&\equiv\sqrt{dt}L_{m}\,\,(1\le m\le N_{C}).\label{eq:Kraus_Vm_def}
\end{align}
Here, $H_\mathrm{eff}$ is the effective (non-Hermitian) Hamiltonian, defined as 
\begin{align}
H_{\mathrm{eff}}\equiv H_S-\frac{i}{2}\sum_m L_m^\dagger L_m.
\label{eq:Heff_definition}
\end{align}
The Kraus operator satisfies the completeness relation $\sum_{m=0}^{N_C} V_m(dt)^\dagger V_m(dt) = \mathbb{I}_S$. 
From the Steinspring representation, $m$ in Eq.~\eqref{eq:Kraus_UM_def}
can be identified as the output when the environment is measured. 
The dynamics conditioned on the output is referred to as the quantum trajectory. 
For detailed information on the continuous measurement, see the recent review paper \cite{Landi:2023:CurFlucReview}.

The matrix product state (MPS) is a mathematical model often used to represent quantum systems comprising multiple particles. Recently, advancements have been made in the use of MPS to account for single-dimensional systems residing in continuous state spaces \cite{Verstraete:2010:cMPS,Osborne:2010:Holography}. 
This development is commonly referred to as cMPS.
The cMPS has proven useful in exploring the thermodynamics of trajectory. It has been applied in the study of phase transitions and the influence of gauge symmetry in both classical and quantum Markov processes \cite{Garrahan:2010:QJ,Lesanovsky:2013:PhaseTrans,Garrahan:2016:cMPS}. Additionally, we recently used cMPS to derive quantum TURs \cite{Hasegawa:2020:QTURPRL,Hasegawa:2021:QTURLEPRL,Hasegawa:2022:FPTTURPRE,Hasegawa:2023:BulkBoundaryBoundNC}.
Consider a cMPS for continuous measurement expressed by \cite{Verstraete:2010:cMPS,Osborne:2010:Holography}
\begin{align}
   \ket{\Psi(\tau)}=\mathcal{V}(\tau)\ket{\psi_{S}(0)}\otimes\ket{\mathrm{vac}}.
   \label{eq:cMPS_state_def}
\end{align}
Here, $\mathcal{V}$ is an operator defined by
\begin{align}
    \mathcal{V}(\tau)=\mathbb{T}e^{-i\int_{0}^{\tau}dt\left(H_{\mathrm{eff}}\otimes\mathbb{I}_F+\sum_{m}iL_{m}\otimes\phi_{m}^{\dagger}(t)\right)},
    \label{eq:Ufrak_def}
\end{align}
where
$\mathbb{T}$ denotes the time-ordering operator,
$\mathbb{I}_F$ is the identity operator in the field, and $\phi_m(s)$ is a field operator that satisfies the canonical commutation relation $[\phi_m(s),\phi_{m^\prime}^\dagger(s^\prime)] =\delta_{mm^\prime} \delta(s-s^\prime)$.
$\ket{\mathrm{vac}}$ is the vacuum state that vanishes with $\phi_m(s)$ for all $m$. 
The cMPS encodes all the information of the continuous measurement by creating particles by applying $\phi_m^\dagger(s)$ to the vacuum state.
The advantage of using the cMPS for continuous measurement is that the
statistics of jump events and the system state can be encoded into a pure state. 
Therefore, we can treat the time evolution of the cMPS as described by closed quantum dynamics. 

First, we review the classical dynamical activity. 
We consider the dynamics within $[0,\tau]$, where $\tau > 0$. 
The dynamical activity within $[0,\tau]$ in a classical Markov process is defined as
\begin{align}
    \mathcal{A}_\mathrm{cl}(\tau)\equiv\int_{0}^{\tau}\sum_{\nu,\mu\,(\nu\ne\mu)}P_{\mu}(t)W_{\nu\mu}(t)dt,
    \label{eq:classical_DA_def}
\end{align}
where $W_{\nu\mu}(t)$ is the transition rate from $\mu$th to $\nu$th state at time $t$ and $P_\mu(t)$ is the probability of being $\mu$th state at time $t$. 
$\mathcal{A}_\mathrm{cl}(t)$ quantifies the average number of jump events within  interval $[0,\tau]$. 
The Lindblad equation describes a classical Markov process by considering $H_S=0$ in Eq.~\eqref{eq:GKSL_eq_def}. 
The classical dynamical activity can then be represented as
\begin{align}
    \mathcal{A}(\tau) \equiv \int_{0}^{\tau}\sum_{m}\mathrm{Tr}_{S}[L_{m}\rho_{S}(t)L_{m}^{\dagger}]dt,
    \label{eq:classical_DA_def2}
\end{align}
where $\mathrm{Tr}_S$ is the partial trace with respect to the system ($\mathrm{Tr}_F$ is defined analogously for the field). 
Equivalence between $\mathcal{A}_\mathrm{cl}(\tau)$ and $\mathcal{A}(\tau)$ for the classical limit can be verified by taking $L_{\nu\mu} = \sqrt{W_{\nu \mu}}\ket{\nu}\bra{\mu}$, where $W_{\nu\mu}$ is the transition rate defined above and $\ket{\mu}$ corresponds to the classical $\mu$th state. 
Let us move on to the quantum dynamical activity. 
Recently, the classical dynamical activity has been generalized to the quantum domain,
which is referred to as the quantum dynamical activity \cite{Hasegawa:2020:QTURPRL,Hasegawa:2023:BulkBoundaryBoundNC}. 
The quantum dynamical activity corresponds to the quantum Fisher information for a particular parametrization of $L_m$ and $H_S$.
Let us first recall the quantum Fisher information in a general scenario. 
Let $\ket{\psi(\vartheta)}$ be an arbitrary state vector parameterized by parameter $\vartheta$. The quantum Fisher information is
\begin{align}
    \mathcal{F}(\vartheta)\equiv4\left[\braket{\partial_{\vartheta}\psi(\vartheta)\mid\partial_{\vartheta}\psi(\vartheta)}-\left|\braket{\partial_{\vartheta}\psi(\vartheta)\mid\psi(\vartheta)}\right|^{2}\right],
    \label{eq:QFI_def}
\end{align}
where $\ket{\partial_{\vartheta}\psi\left(\vartheta\right)}\equiv\partial_{\vartheta}\ket{\psi\left(\vartheta\right)}$. 
Using Eq.~\eqref{eq:QFI_def}, we can introduce the quantum dynamical activity. 
Let $\theta$ be a hypothetical parameter to be estimated, defined as $\theta \equiv t/\tau$. 
For a classical Markov process,
the classical dynamical activity is identical to
the Fisher information of the scaled path probability multiplied by $\theta^2$.
Therefore, the quantum dynamical activity is analogously defined by \cite{Hasegawa:2023:BulkBoundaryBoundNC}
\begin{align}
    \label{eq:QDA_def}
\mathcal{B}(t)=4\theta^2\left(\braket{\partial_{\theta}\Psi(\tau; \theta)|\partial_{\theta}\Psi(\tau; \theta)}-|\braket{\Psi(\tau;\theta)|\partial_{\theta}\Psi(\tau;\theta)}|^2\right),
\end{align}
where $\ket{\Psi(\tau;\theta)}$ is the cMPS parametrized by
\begin{align}
    \ket{\Psi\left(\tau;\theta\right)}\equiv\mathcal{V}\left(\tau,0;\theta\right)\ket{\psi_{S}(0)}\otimes\ket{\mathrm{vac}},\label{eq:cMPS_def_main}
\end{align}
with 
$\mathcal{V}(s_2,s_1;\theta)$
being the operator defined as
\begin{align}
    \mathcal{V}\left(s_{2},s_{1};\theta\right)\equiv\mathbb{T}e^{\int_{s_{1}}^{s_{2}}ds\left(-i\theta H_{\mathrm{eff}}\otimes\mathbb{I}_{\mathrm{F}}+\sqrt{\theta}\sum_{m}L_{m}\otimes\phi_{m}^{\dagger}(s)\right)}.
    \label{eq:ap_psi_main}
\end{align}
For $\theta = 1$, $\ket{\Psi(\tau;\theta)}$ is reduced to the cMPS in Eq.~\eqref{eq:cMPS_state_def} and $\mathcal{B}(t)$ becomes $\mathcal{B}(\tau)$.
Having defined the quantum dynamical activity, we now proceed to its calculation. 
The general Fisher quantum information $\mathcal{F}(\vartheta)$ defined in Eq.~\eqref{eq:QFI_def} can be represented as $\mathcal{F}(\vartheta)=\frac{8}{d\vartheta^{2}}(1-|\langle\psi(\vartheta+d\vartheta)\mid\psi(\vartheta)\rangle|)$. 
Therefore, using Eq.~\eqref{eq:QDA_def},
the quantum dynamical activity can be calculated as
\begin{align}
    \mathcal{B}(t)=\frac{8\theta^{2}}{d\theta^{2}}\left(1-\left|\braket{\Psi(\tau;\theta+d\theta)|\Psi(\tau;\theta)}\right|\right).
    \label{eq:QDA_direct}
\end{align}
Recall that $\braket{\Psi(\tau;\theta+d\theta|\Psi(\tau;\theta))} = \mathrm{Tr}_S[\ket{\Psi(\tau;\theta)}\bra{\Psi(\tau;\theta+d\theta)}] = \mathrm{Tr}_S[\varrho(\tau)]$, 
where $\varrho(\tau)$ obeys the two-sided Lindblad equation \cite{Gammelmark:2014:QCRB}.
We can numerically calculate the quantum dynamical activity directly using a sufficiently small $d\theta$.

In addition to direct numerics, an asymptotic representation of $\mathcal{B}(\tau)$ is also known. 
Suppose that Eq.~\eqref{eq:GKSL_eq_def} has a steady-state solution.
For $\tau \to \infty$, Ref.~\cite{Hasegawa:2020:QTURPRL} showed that
the quantum dynamical activity $\mathcal{B}(\tau)$ can be represented as
\begin{align}
   \mathcal{B}_\mathrm{\infty}(\tau)\equiv \tau\left(\mathfrak{a}+\mathfrak{b}_{c}\right)\;\;\;(\tau \to \infty),
    \label{eq:Bt_long_tau}
\end{align}
where the first term corresponds to the rate of the classical dynamical activity:
\begin{align}
    \mathfrak{a}\equiv \sum_m \mathrm{Tr}_{S}\left[L_{m}\rho_{S}^{\mathrm{ss}}L_{m}^{\dagger}\right],
    \label{eq:At_Tr_def}
\end{align}
with $\rho_S^\mathrm{ss}$ being the steady-state density operator of the Lindblad equation, $\mathcal{L}\rho_S^\mathrm{ss} = 0$. 
Equation~\eqref{eq:At_Tr_def} corresponds to the classical dynamical activity [Eq.~\eqref{eq:classical_DA_def2}]. 
In Eq.~\eqref{eq:Bt_long_tau}, the second term $\mathfrak{b}_c$ quantifies the effect of coherent time evolution in the Lindblad equation 
(see Appendix~\ref{sec:asymp_approach} for the expression).
As mentioned in the classical dynamical activity, $\mathcal{A}(\tau)$ quantifies the extent of the activity of dynamics.
The Lindblad equation consists of two contributions: smooth dynamics induced by the effective Hamiltonian $H_\mathrm{eff}$ and discontinuous dynamics induced by the jump operators $L_m$.
Since $\mathcal{A}(\tau)$ includes only the contribution from $L_m$,
$\mathfrak{b}_c$ reflects the dynamics of the effective Hamiltonian. 
The evaluation of $\mathfrak{b}_c$ requires a pseudo-inverse calculation in the Choi-Jamio{\l}kowski isomorphism that is difficult to perform in general dynamics. 
Equation~\eqref{eq:Bt_long_tau} shows that, for $\tau \to \infty$, the quantum dynamical activity is linear over time. Therefore, at least for $\tau \to \infty$, we cannot expect superlinear scaling of the quantum dynamical activity.

\section{Results}

Although quantum dynamical activity plays an important role in QSL and TUR,
the calculation relies on direct numerics [Eq.~\eqref{eq:QDA_direct}], or asymptotic calculations for $\tau \to \infty$ [Eq.~\eqref{eq:Bt_long_tau}].
In this study, we derive $\mathcal{B}(t)$ analytically and its upper bound of $\mathcal{B}(t)$, which has a clear physical interpretation.
Our first result is the analytical expression of $\mathcal{B}(t)$.
We define the adjoint Lindblad equation for the operator $\mathcal{O}$ as follows:
\begin{align}
    \dot{\mathcal{O}}=\mathcal{L}^{\dagger}\mathcal{O}\equiv i\left[H_{S},\mathcal{O}\right]+\sum_{m=1}^{N_{C}}\mathcal{D}^{\dagger}[L_{m}]\mathcal{O},
    \label{eq:adjoint_Lindblad_def}
\end{align}
where $\mathcal{L}^\dagger$ is the adjoint superoperator
with $\mathcal{D}^\dagger$ being the adjoint dissipator defined by $\mathcal{D}^{\dagger}[L]\mathcal{O}\equiv L^{\dagger}\mathcal{O} L-\frac{1}{2}\left\{ L^{\dagger}L,\mathcal{O}\right\} $. 
The adjoint Lindblad equation is employed for the time evolution of the Hamiltonian or jump operators as opposed to the density operators. This concept aligns with the Heisenberg picture of quantum mechanics.
Subsequently, we find that the exact solution to $\mathcal{B}(\tau)$ is
\begin{widetext}
\begin{align}
    \mathcal{B}(\tau)=\mathcal{A}(\tau)+8\int_{0}^{\tau}ds_{1}\int_{0}^{s_{1}}ds_{2}\mathrm{Re}\left(\mathrm{Tr}_{S}\left[H_{\mathrm{eff}}^{\dagger}\check{H}_{S}\left(s_{1}-s_{2}\right)\rho_{S}\left(s_{2}\right)\right]\right)-4\left(\int_{0}^{\tau}ds\mathrm{Tr}_{S}\left[H_{S}\rho_{S}(s)\right]\right)^{2},
    \label{eq:main_result1}
\end{align}
\end{widetext}
where
$\check{H}_S(t) \equiv e^{\mathcal{L}^\dagger t}H_S$ is the Heisenberg interpretation of the Hamiltonian $H_S$.
For example, in the closed quantum limit $L_m=0$, the operator becomes
$e^{\mathcal{L}^\dagger t}\mathcal{O} = e^{iH_S t}\mathcal{O} e^{-iH_St}$.
Equation~\eqref{eq:main_result1} represents the first result of this study. 
A detailed derivation of Eq.~\eqref{eq:main_result1} is shown in 
Appendix~\ref{sec:QFI_derivation}.
Note that Eq.~\eqref{eq:main_result1} is represented only by the physical quantities of the primary system. 
Equation~\eqref{eq:main_result1} shows that $\mathcal{B}(t)$ comprises the classical contribution $\mathcal{A}(t)$ and quantum correction given by the second and third terms. 
For $\tau \to \infty$, Eq.~\eqref{eq:main_result1} is identical to Eq.~\eqref{eq:Bt_long_tau}
(Appendix~\ref{sec:asymp_approach}). 
We now discuss the limiting cases in Eq.~\eqref{eq:main_result1}. 
For the classical limit, where $H_S=0$, $\mathcal{B}(\tau)$ is reduced to $\mathcal{A}(\tau)$.
By contrast, for the closed quantum limit, where $L_m = 0$, $\mathcal{B}(\tau)$ becomes
\begin{align}
    {\mathcal{B}}(\tau)=4\tau^{2}\left(\mathrm{Tr}_{S}\left[H_{S}^{2}\rho_{S}\right]-\mathrm{Tr}_{S}\left[H_{S}\rho_{S}\right]^{2}\right),
    \label{eq:Bt_quantum_limit}
\end{align}
which is the variance of $H_S$ multiplied by $4\tau^2$. 
Substituting Eq.~\eqref{eq:Bt_quantum_limit} into Eq.~\eqref{eq:QSL_B1} reproduces the Mandelstam-Tamm bound \cite{Mandelstam:1945:QSL}.
This indicates that the exact representation covers the two limiting cases (the classical limit and the closed quantum limit).

After completing our calculations, we realized that Ref.~\cite{Nakajima:2023:SLD} analytically calculated the quantum dynamical activity.
Reference~\cite{Nakajima:2023:SLD} presents the following expression:
\begin{align}
    \mathcal{B}(\tau)=\mathcal{A}(\tau)+4\left(I_{1}+I_{2}\right)-4\ensuremath{\left(\int_{0}^{\tau}ds\mathrm{Tr}_{S}\left[H_{S}(s)\rho_{S}(s)\right]\right)^{2}}.
    \label{eq:Bt_Nakajima}
\end{align}
Here, $I_1$ and $I_2$ are defined as 
$I_1 \equiv \int_0^\tau d s_1 \int_0^{s_1} d s_2 \operatorname{Tr}_S\left[\mathcal{K}_2 \exp \left(\mathcal{L}\left(s_1-s_2\right)\right) \mathcal{K}_1 \rho_S\left(s_2\right)\right]$ and $I_2 \equiv \int_0^\tau d s_1 \int_0^{s_1} d s_2 \operatorname{Tr}_S\left[\mathcal{K}_1 \exp \left(\mathcal{L}\left(s_1-s_2\right)\right) \mathcal{K}_2 \rho_S\left(s_2\right)\right]$, respectively. Here, $\mathcal{K}_1$ and $\mathcal{K}_2$ are superoperators defined by $\mathcal{K}_{1}\bullet=-iH_{\mathrm{eff}} \bullet +\frac12 \sum_k L_k \bullet L_k^{\dagger}$ and $\mathcal{K}_{2}\bullet=i \bullet H_{\mathrm{eff}}^{\dagger}+\frac12 \sum_k L_k \bullet L_k^{\dagger}=(\mathcal{L}-\mathcal{K}_{1})\bullet$, respectively. 
Moreover, Ref.~\cite{Nakajima:2023:SLD} showed that Eq.~\eqref{eq:Bt_Nakajima} can be reduced to Eq.~\eqref{eq:Bt_long_tau} for $\tau \to \infty$. 
However, the expression in Eq.~\eqref{eq:Bt_Nakajima} is represented by superoperators $\mathcal{K}_1$ and $\mathcal{K}_2$, which are difficult to interpret physically.
In Appendix~\ref{sec:from_Nakajima},
we show that our result [Eq.~\eqref{eq:main_result1}] can be derived via Eq.~\eqref{eq:Bt_Nakajima}.

Another advantage of deriving the exact representation is that it is possible to obtain bounds with a more intuitive physical interpretation. 
Specifically, from Eq.~\eqref{eq:main_result1}, we can obtain an upper bound to $\mathcal{B}(\tau)$ comprising the standard deviation of the operators in the Lindblad equation, that is, $\mathcal{B}(\tau) \le \overline{\mathcal{B}}(\tau)$, 
where $\overline{\mathcal{B}}(t)$ denotes its upper bound given by
\begin{align}
    \overline{\mathcal{B}}(\tau)\equiv\mathcal{A}(\tau)+8\int_{0}^{\tau}ds_{1}\sigma_{H_{S}}(s_{1})\int_{0}^{s_{1}}ds_{2}\sigma_{H_{\mathrm{eff}}}(s_{2}),
    \label{eq:BUB_def}
\end{align}
where $\sigma_{\mathcal{O}}(s)\equiv\sqrt{\left\langle \left(\mathcal{O}-\left\langle \mathcal{O}\right\rangle (s)\right)^{\dagger}\left(\mathcal{O}-\left\langle \mathcal{O}\right\rangle (s)\right)\right\rangle }=\sqrt{\left\langle \mathcal{O}^{\dagger}\mathcal{O}\right\rangle (s)-\left|\left\langle \mathcal{O}\right\rangle (s)\right|^{2}}$ for arbitrary operator $\mathcal{O}$.
When $\mathcal{O}$ is Hermitian, $\sigma_\mathcal{O}$ corresponds to the conventional standard deviation. 
Equation~\eqref{eq:BUB_def} is the second result of this study, the proof of which is shown in 
Appendix~\ref{sec:UB_derivation}. 
Some comments are in order. 
The upper bound of Eq.~\eqref{eq:BUB_def} comprises two contributions: the classical dynamical activity $\mathcal{A}(\tau)$ and the quantum correction given by the standard deviations of $H_S$ and $H_\mathrm{eff}$. 
For $\tau\ll 1$, the major contribution comes from the classical part, which indicates that $\mathcal{B}(\tau)$ is linear with respect to $t$ for a short time. 
However, for $\tau \to \infty$, $\mathcal{B}(\tau)$ depends linearly on $t$ when the Lindblad equation has a single steady-state solution \cite{Hasegawa:2020:QTURPRL}. 
Therefore, the upper bound of Eq.~\eqref{eq:BUB_def} becomes loose for a large $\tau$ because the second contribution in Eq.~\eqref{eq:BUB_def} is of $O(\tau^2)$. 
We also comment on the equality condition when $\mathcal{B}(\tau) = \overline{\mathcal{B}}(\tau)$.
Verifying that this equality is satisfied for the classical limit ($H_S=0$) and the closed quantum limit ($L_m=0$) is straightforward. 
As mentioned previously, $\mathcal{B}(\tau)$ is $O(\tau)$ for $\tau \to \infty$ as long as the Lindblad equation has a single steady-state solution while
$\overline{\mathcal{B}}(\tau)$ is $O(\tau^2)$. Therefore it is unlikely that equality is satisfied in other cases (we discuss the equality condition in 
Appendix~\ref{sec:UB_derivation}. 

The upper bound of Eq.~\eqref{eq:BUB_def} includes the standard deviations of $H_S$ and $H_\mathrm{eff}$. 
It is possible to obtain another upper bound that uses the standard deviations of $H_S$ and $\sum_m L_m^\dagger L_m$ 
(please see Eq.~\eqref{eq:ap_ub_result2} in Appendix~\ref{sec:UB_derivation}),
which is not as tight as Eq.~\eqref{eq:BUB_def}.

\begin{figure}[t]
\includegraphics[width=8.7cm]{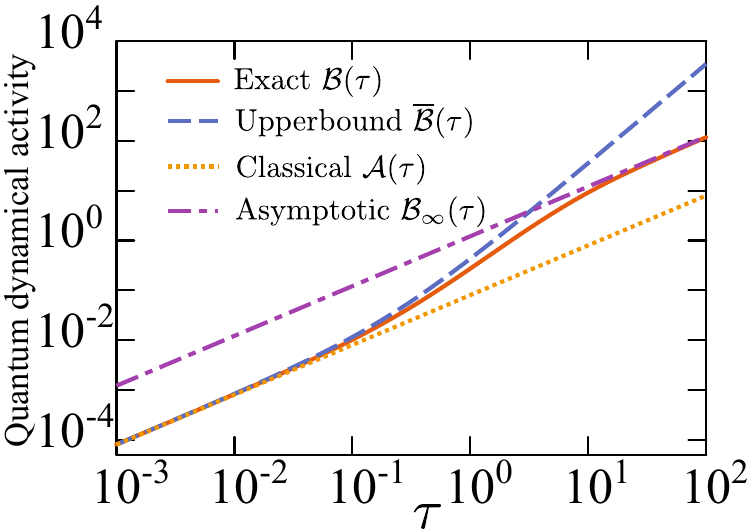} 
\caption{ 
Comparison of quantum dynamical activity as a function of $t$
for the driven two-level model
calculated by the exact method $\mathcal{B}(\tau)$ [Eq.~\eqref{eq:main_result1}] (the solid line) and the asymptotic method $\mathcal{B}_{\infty}(\tau)$ [Eq.~\eqref{eq:Bt_long_tau}] (the dot-dashed line). 
We also provide the corresponding classical dynamical activity $\mathcal{A}(\tau)$ [Eq.~\eqref{eq:classical_DA_def2}] (the dotted line) and the upperbound $\overline{\mathcal{B}}(\tau)$ [Eq.~\eqref{eq:BUB_def}] (the dashed line). 
Since results corresponding to the exact and direct
[Eq.~\eqref{eq:QDA_direct}]
methods completely agree,
we only show the result of the exact method. 
The parameters are $\Omega = 1$, $\kappa = 0.5$, and $\Delta = 1$. 
}
\label{fig:comprison}
\end{figure}

\section{Numerical simulation}

To verify the exact expression [Eq.~\eqref{eq:main_result1}] and the upper bound [Eq.~\eqref{eq:BUB_def}] of $\mathcal{B}(\tau)$, we perform numerical simulations. 
Consider a widely employed two-level atom model driven by a classical laser field. 
Let $\ket{e}$ and $\ket{g}$ be the excited and ground states, respectively. 
The Hamiltonian and jump operators are given by
\begin{align}
H_S&=\Delta\ket{e}\bra{e}+\frac{\Omega}{2}\left(\ket{e}\bra{g}+\ket{g}\bra{e}\right),\label{eq:two_level_Hamiltonian}\\L&=\sqrt{\kappa}\ket{g}\bra{e},\label{eq:two_level_jump}
\end{align}
where $\Delta$, $\Omega$, and $\kappa$ are model parameters.
The jump operator $L$ induces a jump from the excited state to the ground state at the transition rate $\kappa$. 
We compare $\mathcal{B}(\tau)$ calculated using the following approaches under steady-state conditions.
The direct method calculates $\mathcal{B}(\tau)$ by using the quantum Fisher information in Eq.~\eqref{eq:QDA_direct}.
The results of the direct method are treated as the ground truth for the quantum dynamical activity. 
The exact method, which is the approach proposed in this study, calculates the quantum dynamical activity using Eq.~\eqref{eq:main_result1}. 
The asymptotic method is based on Eq.~\eqref{eq:Bt_long_tau},
which holds for $\tau \to \infty$. 
We also evaluate the upper bound $\overline{\mathcal{B}}(\tau)$ defined by Eq.~\eqref{eq:BUB_def}. 
For comparison, we calculate the classical dynamical activity using Eq.~\eqref{eq:At_Tr_def}.

Figure~\ref{fig:comprison} shows the comparison of the methods.
We plot the quantum dynamical activity as a function of $\tau$. 
First, we compare the direct method [Eq.~\eqref{eq:QDA_direct}] and the exact solution [Eq.~\eqref{eq:main_result1}] to determine whether the two approaches agree completely.
This verifies the validity of Eq.~\eqref{eq:main_result1}.
Therefore, we do not plot the direct method, as shown in Fig.~\ref{fig:comprison}. 
In Fig.~\ref{fig:comprison}, the solid, dot-dashed, and dotted lines denote the exact solution, the asymptotic solution, and the classical dynamical activity, respectively. 
The asymptotic method converges to the direct method for $\tau \to \infty$. However, the two results are approximately 10 times apart for a smaller $\tau$, indicating that the exact method presented in this manuscript is important for a short time $\tau$. 
The exact result for $\tau \to 0$ is close to that for the classical quantum dynamical activity,
indicating that quantumness cannot improve the precision of the counting observable for a short time according to the quantum TUR [Eq.~\eqref{eq:QTUR1}]. 
For the two limiting cases $\tau \to 0$ and $\tau \to \infty$, the quantum dynamical activity is linear with respect to time. 
In the intermediate interval, namely from $\tau \sim 10^{-1}$ to $\tau = 10^1$,
the quantum dynamical activity exhibits superlinear scaling with respect to time. 
This implies that the precision of the counting observable can be improved superlinearly. 
This behavior broadly applies to general dynamics, as established by the upper bound in Eq.~\eqref{eq:BUB_def}. 
The upper bound of the quantum dynamical activity scales as $O(\tau) + O(\tau^2)$.
Therefore, when $\tau \ll 1$, the $O(\tau)$ component dominates the quantum dynamical activity, indicating that we cannot expect a quantum advantage over a short time period. 
Next, we consider the upper bound derived from Eq.~\eqref{eq:BUB_def},
plotted using the dashed line in Fig.~\ref{fig:comprison}.
The upper bound is above the exact solution (solid line),
which verifies the accuracy of the upper bound. 
Because the quantum dynamical activity is linear as a function of $\tau$ for $\tau\to \infty$, the upper bound becomes loose for this large time limit, because the upper bound is $O(\tau^2)$. 

Thus far, we have discussed the effects of the difference between classical and quantum dynamical activities on the TUR.
When considering the TUR, the lower bound is given by the reciprocal of the dynamical activity, allowing a simple understanding of the qualitative difference between linear and superlinear scaling with respect to $\tau$. However, this clarity disappears with the speed limits, because their upper bounds include the integral (right-hand side of Eq.~\eqref{eq:QSL_B1}), which obscures clear differences. Nevertheless, a quantum system provides a larger proportionality constant, resulting in a higher upper bound.

\section{Conclusion}

In this study, we derived the exact solution for the quantum dynamical activity using cMPS. In addition, we obtained the upper bound for the quantum dynamical activity, which comprises the standard deviation of the operators of the Lindblad equation.
Numerical simulations are performed to validate the results. Our findings are expected to enhance our understanding of quantum nonequilibrium dynamics, considering the crucial role of quantum dynamical activity in uncertainty relations, such as QSLs and quantum TURs. 

\begin{acknowledgments}
This work was supported by JSPS KAKENHI Grant Number JP22H03659.
\end{acknowledgments}

\begin{widetext}

\appendix

\section{Derivation of exact solution [Eq.~\eqref{eq:main_result1}]\label{sec:QFI_derivation}}
\subsection{Preparation}
We derive ancillary relations to derive Eq.~\eqref{eq:main_result1}.
Let us apply $\phi_m(s')$ and $\phi_m(s')^\dagger$ at time $s'\in[0,s)$ to $\ket{\psi_{S}(0)}\otimes\ket{\mathrm{vac}}\bra{\psi_{S}(0)}\otimes\bra{\mathrm{vac}}$ from left and right, and denote by the resulting operator $\mathcal{F}\left([0,s)\right)$. 
For $t\geq s$, the first relations are given by
    \begin{align}
        \label{eq:ap_phi}
        &\mathrm{Tr}
        _{F,S}\left[d\phi_m(t)\mathcal{F}\left([0,s)\right)\right]=\mathrm{Tr}
        _{F,S}\left[d\phi^\dagger_m(t)\mathcal{F}\left([0,s)\right)\right]=0,
    \end{align}
    \begin{align}
    \label{eq:ap_phiphi}
        &\mathrm{Tr}
        _{F,S}\left[d\phi_m(t) d\phi_{m'}^{\dagger}(t)\mathcal{F}\left([0,s)\right)\right]=ds\delta_{m,m'}\mathrm{Tr}
        _{F,S}\left[\mathcal{F}\left([0,s)\right)\right],
    \end{align}
where
\begin{align}
    d\phi_m(s)\equiv \int_s^{s+ds} ds' \phi_m(s'),
\end{align}
and $\mathrm{Tr}_{F,S}[\bullet]$ denotes $\mathrm{Tr}_{S}\left[\mathrm{Tr}_{F}\left[\bullet\right]\right]=\mathrm{Tr}_{F}\left[\mathrm{Tr}_{S}\left[\bullet\right]\right]$.
Equations~\eqref{eq:ap_phi} and~\eqref{eq:ap_phiphi} can be derived from the canonical commutation relation. 
The second relation is 
the unitarity of $\mathcal{V}$:
    \begin{align}
        \label{eq:ap_VV}
        &\mathrm{Tr}
        _{F,S}\left[ \mathcal{V}\left(s, s_1; \theta\right)\mathcal{F}\left([0,s_1)\right)\mathcal{V}\left(s, s_1; \theta\right)^\dagger\right]=\mathrm{Tr}
        _{F,S}\left[\mathcal{F}\left([0,s_1)\right)\right].
    \end{align}
Using $\mathcal{V}\left(s, s_1; \theta\right)=\mathcal{V}\left(s, s-ds; \theta\right)\mathcal{V}\left(s-ds, s_1; \theta\right)$ and the cyclic property of the trace, we have
    \begin{align}
        \label{eq:ap_VV_d}
        &\mathrm{Tr}
        _{F,S}\left[ \mathcal{V}\left(s, s-ds; \theta\right)^\dagger \mathcal{V}\left(s, s-ds; \theta\right)\mathcal{V}\left(s-ds, s_1; \theta\right)\mathcal{F}\left([0,s_1)\right)\mathcal{V}\left(s-ds, s_1; \theta\right)^\dagger\right] \nonumber\\
        &=\mathrm{Tr}
        _{F,S}\left[ \mathcal{V}\left(s-ds, s_1; \theta\right)\mathcal{F}\left([0,s_1)\right) \mathcal{V}\left(s-ds, s_1; \theta\right)^\dagger\right]+O(ds^2),
    \end{align}
where we use Eq.~\eqref{eq:ap_phi} and Eq.~\eqref{eq:ap_phiphi}.
By repeatedly using Eq.~\eqref{eq:ap_VV_d}, we have Eq.~\eqref{eq:ap_VV}.
Since $\mathrm{Tr}_{F}\left[ \bra{\Psi\left(\tau;\theta\right)}\ket{\Psi\left(\tau;\theta\right)}\right]$ provides the density matrix of the time scaled by $\theta$ (Ref.~\cite{Hasegawa:2023:BulkBoundaryBoundNC}), the third relation is given by 
    \begin{align}
        \label{eq:ap_rho_VV}
        \mathrm{Tr}_{F}\left[\ket{\Psi(s;\theta)} \bra{\Psi(s;\theta)} \right]=\rho_S\left(\theta s\right).
    \end{align}
\subsection{Derivation}
By differentiating Eq.~\eqref{eq:cMPS_def_main} with respect to $\theta$, we obtain 
    \begin{align}
        \label{eq:ap_diff}
        &\ket{\partial_{\theta}\Psi\left(\tau;\theta\right)}=\int_0^{\tau-ds}  \mathcal{V}\left(\tau, s+ds; \theta\right)d\mathcal{V}\left(s; \theta\right)\mathcal{V}\left(s, 0; \theta\right)\ket{\psi_{S}(0)} \otimes\ket{\mathrm{vac}}=\int_0^{\tau-ds}  \mathcal{V}\left(\tau, s+ds; \theta\right)d\mathcal{V}\left(s; \theta\right)\ket{\Psi(s;\theta)},
    \end{align}
where 
\begin{align}
    \label{eq:ap_defdV}
    d\mathcal{V}\left(s; \theta\right)\equiv \partial_{\theta}\mathcal{V}\left(s+ds, s; \theta\right)= -i H_{\mathrm{eff}}\otimes \mathbb{I}_F ds+\frac{1}{2\sqrt{\theta}}\sum_m L_m \otimes d\phi_m^\dagger(s).
\end{align}
First, we calculate the term 
\begin{align}
    \label{eq:ap_tr_psi}    \mathcal{K}\equiv\braket{\partial_{\theta}\Psi\left(\tau;\theta\right)|\partial_{\theta}\Psi\left(\tau;\theta\right)}=\mathrm{Tr}_{F,S}\left[\ket{\partial_{\theta}\Psi\left(\tau;
    \theta\right)}\bra{\partial_{\theta}\Psi\left(\tau;\theta\right)}\right].
\end{align}
Substituting Eq.~\eqref{eq:ap_diff} into this relation, we obtain
    \begin{align}
        \label{eq:ap_K}
        &\mathcal{K}=\int_0^{\tau-ds} \int_0^{\tau-ds} \mathrm{Tr}_{F,S}\left[ \mathcal{V}\left(\tau, s_1+ds; \theta\right)d\mathcal{V}\left(s_1; \theta\right)\ket{\Psi(s_1;\theta )} 
        \bra{\Psi(s_2;\theta )}  d\mathcal{V}\left(s_2; \theta\right)^\dagger\mathcal{V}\left(\tau, s_2+ds; \theta\right)^\dagger\right].
    \end{align}
Here, the first integral corresponds to $s_1$ and the second integral corresponds to $s_2$.
We decompose $\mathcal{K}$ into the sum of terms $s_1=s_2$ and $s_1\neq s_2$ which are written as $\mathcal{K}_{s_1=s_2}$ and $\mathcal{K}_{s_1\neq s_2}$.
When $s_1=s_2=s$, from Eq.~\eqref{eq:ap_VV}, we obtain
    \begin{align}
        &\mathcal{K}_{s_1=s_2}=\int_0^{\tau-ds} \mathrm{Tr}_{F,S}\left[d\mathcal{V}\left(s; \theta\right)\ket{\Psi(s;\theta )} \bra{\Psi(s;\theta )} d\mathcal{V}\left(s; \theta\right)^\dagger\right]=\int_0^{\tau-ds} \mathrm{Tr}_{F,S}\left[d\mathcal{V}\left(s; \theta\right)^\dagger d\mathcal{V}\left(s; \theta\right)\ket{\Psi(s;\theta)} \bra{\Psi(s;\theta)} \right].
    \end{align}
By using Eq.~\eqref{eq:ap_phi}, Eq.~\eqref{eq:ap_phiphi} for 
$d\mathcal{V}\left(s; \theta\right)^\dagger d\mathcal{V}\left(s; \theta\right)$, and using Eq.~\eqref{eq:ap_rho_VV}, we obtain
    \begin{align}
        \label{eq:ap_K_eq_result}
        \mathcal{K}_{s_1=s_2}&=\frac{1}{4\theta}\int_0^{\tau-ds} ds \mathrm{Tr}_{F,S}\left[\sum_m L_m^\dagger L_m \ket{\Psi(s;\theta)} \bra{\Psi(s;\theta)} \right]=\frac{1}{4\theta^2}\int_0^t ds\sum_m \mathrm{Tr}_{S}\left[ L_m \rho_S(s) L_m^\dagger\right]=\frac{1}{4\theta^2}\mathcal{A}(t).
    \end{align}
Next, we calculate $\mathcal{K}_{s_1\neq s_2}$.
When $s_1>s_2$, applying Eq.~\eqref{eq:ap_VV} to Eq.~\eqref{eq:ap_K}, we obtain
    \begin{align}
        \int_0^{\tau-ds} \int_0^{s_1-ds} \mathrm{Tr}_{F,S}\left[d\mathcal{V}\left(s_1; \theta\right)\ket{\Psi(s_1;\theta )} \bra{\Psi( s_2;\theta)} d\mathcal{V}\left(s_2; \theta\right)^\dagger\mathcal{V}\left(s_1, s_2+ds; \theta\right)^\dagger \mathcal{V}\left(s_1+ds, s_1; \theta\right)^\dagger\right].
    \end{align}
From $\mathrm{Tr}[\mathcal{F}\mathcal{G}]^*=\mathrm{Tr}[\mathcal{G}^{\dagger}\mathcal{F}^{\dagger}]$, we find that the case of $s_1<s_2$ is the complex conjugate of this equation. Therefore, using the cyclic property of the trace, we obtain
    \begin{align}
        \label{eq:ap_noeq}
        \mathcal{K}_{s_{1}\neq s_{2}}&=2\int_{0}^{\tau-ds}\int_{0}^{s_{1}-ds}\mathrm{Re}\left(\mathrm{Tr}_{F,S}\left[\mathcal{V}\left(s_{1}+ds,s_{1};\theta\right)^{\dagger}d\mathcal{V}\left(s_{1}; \theta\right)\ket{\Psi(s_1;\theta )} \bra{\Psi(s_2;\theta )}d\mathcal{V}\left(s_{2}; \theta\right)^{\dagger}\mathcal{V}\left(s_{1},s_{2}+ds;\theta\right)^{\dagger}\right]\right).
    \end{align}
Applying Eq.~\eqref{eq:ap_phi} and Eq.~\eqref{eq:ap_phiphi} for $ \mathcal{V}\left(s_1+ds, s_1; \theta\right)^\dagger d\mathcal{V}\left(s_1; \theta\right)$, we obtain
    \begin{align}
        \label{eq:ap_noeq2}
        &\mathcal{K}_{s_{1}\neq s_{2}}=2\int_{0}^{\tau}ds_{1}\int_{0}^{s_{1}-ds}\mathrm{Re}\left(\mathrm{Tr}_{F,S}\left[\left(-i H_{\mathrm{eff}}+\frac{1}{2}\sum_{m}L_{m}^{\dagger}L_{m}\right)\ket{\Psi(s_1;\theta )} \bra{\Psi(s_2;\theta )}\mathcal{V}\left(s_{2}; \theta\right)^{\dagger}\mathcal{V}\left(s_{1},s_{2}+ds;\theta\right)^{\dagger}\right]\right)\nonumber\\&=-2\int_{0}^{\tau}ds_{1}\int_{0}^{s_{1}-ds}\mathrm{Re}\left(\mathrm{Tr}_{F,S}\left[iH_{S}\ket{\Psi(s_1;\theta )} \bra{\Psi(s_2;\theta )}d\mathcal{V}\left(s_{2}; \theta\right)^{\dagger}\mathcal{V}\left(s_{1},s_{2}+ds;\theta\right)^{\dagger}\right]\right).
    \end{align}
Considering the term proportional to $d\phi_m(s_2)$ in $d\mathcal{V}\left(s_{2}; \theta\right)^{\dagger}$ in Eq.~\eqref{eq:ap_noeq2}, we obtain
    \begin{align}
        \label{eq:ap_noeq3}
        -\int_{0}^{\tau}ds_{1}\int_{0}^{s_{1}-ds}\mathrm{Re}\left(\mathrm{Tr}_{F,S}\left[iH_{S}\ket{\Psi(s_1;\theta )} \bra{\Psi(s_2;\theta )}\frac{1}{\sqrt{\theta}}\sum_m L_m^{\dagger} \otimes d\phi_m(s_2)\mathcal{V}\left(s_{1},s_{2}+ds;\theta\right)^{\dagger}\right]\right).
    \end{align}
Using the canonical commutation relation and combining $d\phi_m(s_2)$ with $d\phi_m^{\dagger}(s_2)$ in
\begin{align}
    \mathcal{V}\left(s_{1},0;\theta\right)=\mathcal{V}\left(s_{1},s_2+ds;\theta\right)\mathcal{V}\left(s_{2}+ds,s_2;\theta\right)\mathcal{V}\left(s_{2},0;\theta\right),
\end{align}
and recall that Eq.~\eqref{eq:cMPS_def_main},
Eq.~\eqref{eq:ap_noeq3} yields
    \begin{align}
        \label{eq:ap_noeq4}
        & -\sum_m\int_{0}^{\tau}ds_{1}\int_{0}^{s_{1}}ds_2\mathrm{Re}\left(\mathrm{Tr}_{F,S}\left[i H_{S}\mathcal{V}\left(s_{1},s_2+ds;\theta\right)L_m\ket{\Psi(s_2;\theta )} \bra{\Psi(s_2;\theta )}L_m^{\dagger}\mathcal{V}\left(s_{1},s_{2}+ds;\theta\right)^{\dagger}\right]\right)\nonumber\\
        &\equiv -\sum_m\int_{0}^{\tau}ds_{1}\int_{0}^{s_{1}}ds_2\mathrm{Re}\left(F_m\left(s_{1},s_{2};\theta\right)\right).
    \end{align}
Since $\mathrm{Tr}\left[\mathcal{F}\right]^*=\mathrm{Tr}\left[\mathcal{F}^{\dagger}\right]$, we have
    \begin{align}
        &F\left(s_{1},s_{2};\theta\right)^*=-\mathrm{Tr}_{F,S}\left[i\mathcal{V}\left(s_{1},s_2+ds;\theta\right)L_m \ket{\Psi(s_2;\theta )} \bra{\Psi(s_2;\theta )}L_m^{\dagger}\mathcal{V}\left(s_{1},s_{2}+ds;\theta\right)^{\dagger}H_{S}\right] =-F\left(s_{1},s_{2};\theta\right),
    \end{align}
where we use the cyclic property of the trace in the second equality.
From this result, we find that $F\left(s_{1},s_{2};\theta\right)$ is a purely imaginary number and Eq.~\eqref{eq:ap_noeq4} is equal to zero.
Therefore, from Eq.~\eqref{eq:ap_defdV} and Eq.~\eqref{eq:ap_noeq2}, we obtain
    \begin{align}
        \label{eq:ap_noneq_eff}
        \mathcal{K}_{s_{1}\neq s_{2}}&=2\int_{0}^{\tau}ds_{1}\int_{0}^{s_{1}}ds_{2}\mathrm{Re}\left(\mathrm{Tr}_{F,S}\left[H_{S}\ket{\Psi(s_1;\theta )} \bra{\Psi(s_2;\theta )}H_{\mathrm{eff}}^{\dagger}\mathcal{V}\left(s_{1},s_{2};\theta\right)^{\dagger}\right]\right).
    \end{align}
From $\mathcal{V}\left(s_{1},0;\theta\right)=\mathcal{V}\left(s_{1},s_{2};\theta\right)\mathcal{V}\left(s_{2},0;\theta\right)$ and Eq.~\eqref{eq:cMPS_def_main}, we obtain
    \begin{align}
        \label{eq:ap2_noneq}
        \mathcal{K}_{s_{1}\neq s_{2}}&
        =2\int_{0}^{\tau}ds_{1}\int_{0}^{s_{1}}ds_{2}\mathrm{Re}\left(\mathrm{Tr}_{S}\left[\mathrm{Tr}_F\left[H_{S}\mathcal{V}\left(s_{1},s_{2};\theta\right)\ket{\Psi(s_2;\theta )} \bra{\Psi(s_2;\theta )}H_{\mathrm{eff}}^{\dagger}\mathcal{V}\left(s_{1},s_{2};\theta\right)^{\dagger}\right]\right]\right)\nonumber\\
        &=2\int_{0}^{\tau}ds_{1}\int_{0}^{s_{1}}ds_{2}\mathrm{Re}\left(\mathrm{Tr}_{S}\left[\mathrm{Tr}_F\left[H_{\mathrm{eff}}^{\dagger}\mathcal{V}\left(s_{1},s_{2};\theta\right)^{\dagger}H_{S}\mathcal{V}\left(s_{1},s_{2};\theta\right)\ket{\Psi(s_2;\theta )} \bra{\Psi(s_2;\theta )}\right]\right]\right).
    \end{align}
Since Eq.~\eqref{eq:ap_phi} and Eq.~\eqref{eq:ap_phiphi} also hold for $\mathrm{Tr}_F[\bullet]$ instead of $\mathrm{Tr}_{F,S}[\bullet]$, we apply these relations to the time in $[s_{1}-ds, s_{1})$:
    \begin{align}
        \label{eq:ap2_tr}
        &\mathrm{Tr}_{F}\left[H_{\mathrm{eff}}^{\dagger}\mathcal{V}\left(s_{1},s_{2};\theta\right)^{\dagger}H_{S}\mathcal{V}\left(s_{1},s_{2};\theta\right)\ket{\Psi(s_2;\theta )} \bra{\Psi(s_2;\theta)}\right]\nonumber\\
        &=H_{\mathrm{eff}}^{\dagger}\mathrm{Tr}_{F}\left[\mathcal{V}\left(s_{1}-ds,s_{2};\theta\right)^{\dagger}\sum_m V_m\left(ds'\right)^ {\dagger}H_{S}V_m\left(ds'\right)\mathcal{V}\left(s_{1}-ds,s_{2};\theta\right)\ket{\Psi(s_2;\theta )} \bra{\Psi(s_2;\theta)}\right],
    \end{align}
where $ds'\equiv \theta ds$, and $V_m$ comprises the Kraus operators, 
which are defined in Eq.~\eqref{eq:Kraus_V0_def} and Eq.~\eqref{eq:Kraus_Vm_def}.
Generalizing Eq.~\eqref{eq:ap2_tr}, we obtain
    \begin{align}
        \label{eq:ap2_tr2}
        &\mathrm{Tr}_{F}\left[H_{\mathrm{eff}}^{\dagger}\mathcal{V}\left(s_{1},s_{2};\theta\right)^{\dagger}H_{S}\mathcal{V}\left(s_{1},s_{2};\theta\right)\ket{\Psi(s_2;\theta )} \bra{\Psi(s_2;\theta)}\right]\nonumber\\
        &=H_{\mathrm{eff}}^{\dagger}\sum_{m_{N-1}}\cdots \sum_{m_0}V_{m_0}(ds')^{\dagger}\cdots V_{m_{N-1}}(ds')^{\dagger} H_{S}V_{m_{N-1}}(ds')\cdots V_{m_0}(ds')\mathrm{Tr}_{F}\left[\ket{\Psi(s_2;\theta )} \bra{\Psi(s_2;\theta)}\right]\nonumber\\
        &=H_{\mathrm{eff}}^{\dagger}\sum_{m_{N-1}}\cdots \sum_{m_0}V_{m_{0}}(ds')^{\dagger}\cdots V_{m_{N-1}}(ds')^{\dagger} H_{S}V_{m_{N-1}}(ds')\cdots V_{m_{0}}(ds')\rho_S\left(\theta s_{2}\right),
    \end{align}
where $N\equiv (s_{1}-s_{2})/ds=\theta(s_{1}-s_{2})/ds'$ and we use Eq.~\eqref{eq:ap_rho_VV} in the last equality.
Since the integration range in Eq.~\eqref{eq:ap2_tr2} is the same as $[0, \theta(s_{1}-s_{2})]=[0, Nds']$, we find that  
    \begin{align}
    \label{eq:ap2_Hhat}
        \check{H}_{S}\left(\theta(s_{1}-s_{2})\right)\equiv \sum_{m_{N-1}}\cdots \sum_{m_0}V_{m_{0}}(ds')^{\dagger}\cdots V_{m_{N-1}}(ds')^{\dagger} H_{S}V_{m_{N-1}}(ds')\cdots V_{m_{0}}(ds')=e^{\mathcal{L}^\dagger \theta(s_1-s_2)}H_S.
    \end{align}
Substituting Eq.~\eqref{eq:ap2_tr2} and Eq.~\eqref{eq:ap2_Hhat} into Eq.~\eqref{eq:ap2_noneq}, we obtain
\begin{align}
    \label{eq:ap2_noneq_result}
    \mathcal{K}_{s_{1}\neq s_{2}}&=2\int_{0}^{\tau}ds_{1}\int_{0}^{s_{1}}ds_{2}\mathrm{Re}\left(\mathrm{Tr}_S\left[H_{\mathrm{eff}}^{\dagger}\check{H}_{S}\left(\theta(s_{1}-s_{2})\right)\rho_S\left(\theta s_{2}\right)\right]\right)=\frac{2}{\theta^2}\int_{0}^{t}ds_{1}\int_{0}^{s_{1}}ds_{2}\mathrm{Re}\left(\mathrm{Tr}_S\left[H_{\mathrm{eff}}^{\dagger}\check{H}_{S}\left(s_{1}-s_{2}\right)\rho_S\left(s_{2}\right)\right]\right).
\end{align}
By combining Eq.~\eqref{eq:ap_K_eq_result} and Eq.~\eqref{eq:ap2_noneq_result}, we obtain
\begin{align}
    \label{eq:ap_noneq_result}
    4\theta^{2}\braket{\partial_{\theta}\Psi\left(\tau;\theta\right)|\partial_{\theta}\Psi\left(\tau;\theta\right)}=\mathcal{A}(t)+8\int_{0}^{t}ds_{1}\int_{0}^{s_{1}}ds_{2}\mathrm{Re}\left(\mathrm{Tr}_{S}\left[H_{\mathrm{eff}}^{\dagger}\check{H}_{S}\left(s_{1}-s_{2}\right)\rho_{S}\left(s_{2}\right)\right]\right).
\end{align}
Finally, we calculate the term $\braket{\Psi\left(\tau;\theta\right)|\partial_{\theta}\Psi\left(\tau;\theta\right)}$. Substituting Eq.~\eqref{eq:ap_diff} into 
\begin{align}
    \braket{\Psi\left(\tau;\theta\right)|\partial_{\theta}\Psi\left(\tau;\theta\right)}=\mathrm{Tr}_{F,S}\left[\ket{\partial_{\theta}\Psi\left(\tau;\theta\right)}\bra{\Psi\left(\tau;\theta\right)}\right],
\end{align}
and using Eq.~\eqref{eq:ap_VV}, we obtain 
    \begin{align}
        \label{eq:ap_psi_dif}
        &\mathrm{Tr}_{F,S}\left[\ket{\partial_{\theta} \Psi\left(\tau;\theta\right)}\bra{\Psi\left(\tau;\theta\right)}\right] =\int_0^{\tau-ds}  \mathrm{Tr}_{F,S}\left[\mathcal{V}\left(\tau, s+ds; \theta\right)d\mathcal{V}\left(s; \theta\right)\ket{\Psi(s;\theta )} \bra{\Psi(s;\theta)}\mathcal{V}\left(s+ds, s; \theta\right)^\dagger\mathcal{V}\left(\tau, s+ds; \theta\right)^\dagger\right]\nonumber\\
        &=\int_0^{\tau-ds} \mathrm{Tr}_{F,S}\left[ d\mathcal{V}\left(s; \theta\right)\ket{\Psi(s;\theta )} \bra{\Psi(s;\theta)}\mathcal{V}\left(s+ds, s; \theta\right)^{\dagger}\right]=\int_0^{\tau-ds} \mathrm{Tr}_{F,S}\left[ \mathcal{V}\left(s+ds, s; \theta\right)^{\dagger}d\mathcal{V}\left(s; \theta\right)\ket{\Psi(s;\theta )} \bra{\Psi(s;\theta)}\right].
    \end{align}
Similarly to Eq.~\eqref{eq:ap_noeq2}, by using Eq.~\eqref{eq:ap_phi}, Eq.~\eqref{eq:ap_phiphi} for $\mathcal{V}\left(s+ds, s; \theta\right)^{\dagger}d\mathcal{V}\left(s; \theta\right)$, we obtain
    \begin{align}
        \label{eq:ap_psi_dif2}
        &\braket{\Psi\left(\tau;\theta\right)|\partial_{\theta}\Psi\left(\tau;\theta\right)}=-i\int_0^{\tau}ds \mathrm{Tr}_{F,S}\left[ H_S\ket{\Psi(s;\theta )} \bra{\Psi(s;\theta)}\right]
        =-\frac{i}{\theta}\int_0^{t}ds \mathrm{Tr}_{S}\left[ H_S\rho_{S} \left(s\right)\right].
    \end{align}
Therefore, we have
\begin{align}
    \label{eq:ap_psi_dif_result}
    4\theta^2|\braket{\Psi\left(\tau;\theta\right)|\partial_{\theta}\Psi\left(\tau;\theta\right)}|^2=4\left(\int_0^{t}ds \mathrm{Tr}_{S}\left[ H_S\rho_{S} \left(s\right)\right]\right)^2.
\end{align}
By substituting Eq.~\eqref{eq:ap_noneq_result} and Eq.~\eqref{eq:ap_psi_dif_result} into Eq.~\eqref{eq:QDA_def}, we obtain Eq.~\eqref{eq:main_result1}.
\section{Derivation of upper bound [Eq.~\eqref{eq:BUB_def}]\label{sec:UB_derivation}}
Let $F$ be an arbitrary operator of the primary system.
We define the mean and standard deviation of $F$ as follows:
\begin{align}
    \label{eq:ap2_def_mean}
    \langle F\rangle(s) &\equiv \mathrm{Tr}_S\left[F \rho_S(s)\right],\\
    \label{eq:ap2_def_stdev}
    \sigma_{F}(s)&\equiv \sqrt{\left\langle \left(F-\left\langle F\right\rangle(s)\mathbb{I}_S\right)^{\dagger}\left(F-\left\langle F\right\rangle(s)\mathbb{I}_S\right) \right\rangle}=\sqrt{\left\langle F^{\dagger} F \right\rangle(s) - \left|\left\langle F\right\rangle(s)\right|^2}.
\end{align}
In the following, we drop $\mathbb{I}_S$ which is multiplied by constants.
We derive the upper bound of $\mathcal{K}_{s_{1}\neq s_{2}}$.
By setting $\theta=1$ in Eq.~\eqref{eq:ap2_tr2} and Eq.~\eqref{eq:ap2_Hhat}, we obtain
    \begin{align}
        \label{eq:ap2_ub1}
         \mathcal{K}_{s_{1}\neq s_{2}}&=\frac{2}{\theta^2}\int_{0}^{t}ds_{1}\int_{0}^{s_{1}}ds_{2}\mathrm{Re}\left(\mathrm{Tr}_S\left[H_{\mathrm{eff}}^{\dagger}\check{H}_{S}\left(s_{1}-s_{2}\right)\rho_S\left(s_{2}\right)\right]\right)\nonumber\\
        &=\frac{2}{\theta^2}\int_{0}^{t}ds_{1}\int_{0}^{s_{1}}ds_{2}\mathrm{Re}\left(\mathrm{Tr}_S\left[\mathrm{Tr}_{F}\left[H_{\mathrm{eff}}^{\dagger}\mathcal{V}\left(s_{1},s_{2}\right)^{\dagger}H_{S}\mathcal{V}\left(s_{1},s_{2}\right)\ket{\Psi(s_2;\theta )} \bra{\Psi(s_2;\theta )}\right]\right]\right)\nonumber\\
        &=\frac{2}{\theta^2}\int_{0}^{t}ds_{1}\int_{0}^{s_{1}}ds_{2}\mathrm{Re}\left(\mathrm{Tr}_{F,S}\left[H_{S}\ket{\Psi(s_1;\theta )} \bra{\Psi(s_2;\theta )}H_{\mathrm{eff}}^{\dagger}\mathcal{V}\left(s_{1},s_{2}\right)^{\dagger}\right]\right),
    \end{align}
where $\mathcal{V}\left(s_{1},s_{2}\right)\equiv \mathcal{V}\left(s_{1},s_{2};1\right)$.
By using Eq.~\eqref{eq:ap_VV} and Eq.~\eqref{eq:ap_rho_VV}, the trace term can be written as
    \begin{align}
        &\mathrm{Tr}_{F,S}\left[H_{S}\ket{\Psi(s_1;\theta )} \bra{\Psi(s_2;\theta )}H_{\mathrm{eff}}^{\dagger}\mathcal{V}\left(s_{1},s_{2}\right)^{\dagger}\right]\nonumber\\ &=\mathrm{Tr}_{F,S}\left[\left(H_{S}-\langle H_{S}\rangle(s_{1})\right)\ket{\Psi(s_1;\theta )} \bra{\Psi(s_2;\theta )}\left(H_{\mathrm{eff}}-\langle H_{\mathrm{eff}}\rangle(s_{2})\right)^{\dagger}\mathcal{V}\left(s_{1},s_{2}\right)^{\dagger}\right]+\langle H_{S}\rangle(s_{1})\langle H_{\mathrm{eff}}\rangle(s_{2})^*.
        \label{eq:ap2_tr_expansion}
    \end{align}
Therefore, we have
    \begin{align}
        \label{eq:ap2_corr}
         \mathcal{K}_{s_{1}\neq s_{2}}&\le \frac{2}{\theta^2}\int_{0}^{t}ds_{1}\int_{0}^{s_{1}}ds_{2}\left|\mathrm{Tr}_{F,S}\left[\Delta H_{S}(s_{1})\ket{\Psi(s_1;\theta )} \bra{\Psi(s_2;\theta )}\Delta H_{\mathrm{eff}}(s_{2})^{\dagger}\mathcal{V}\left(s_{1},s_{2}\right)^{\dagger}\right]\right|\nonumber\\
         &+\frac{2}{\theta^2}\int_{0}^{t}ds_{1}\int_{0}^{s_{1}}ds_{2}\mathrm{Re}\left(\langle H_{S}\rangle(s_{1})\langle H_{\mathrm{eff}}\rangle(s_{2})^*\right),
    \end{align}
where $\Delta F(s)\equiv F-\langle F \rangle(s)$.
The second term on the right-hand side can be written as
\begin{align}
    \label{eq:ap2_second}
    \frac{2}{\theta^2}\int_{0}^{t}ds_{1}\int_{0}^{s_{1}}ds_{2}\mathrm{Re}\left(\langle H_{S}\rangle(s_{1})\langle H_{\mathrm{eff}}\rangle(s_{2})^*\right)=\frac{2}{\theta^2}\int_{0}^{t}ds_{1}\int_{0}^{s_{1}}ds_{2}\langle H_{S}\rangle(s_{1})\langle H_{S}\rangle(s_{2})=\frac{1}{\theta^2}\left(\int_{0}^{t}ds \langle H_{S}\rangle(s)\right)^2.
\end{align}
Regarding the first term on the right-hand side in Eq.~\eqref{eq:ap2_corr}, by applying the Cauchy-Schwarz inequality, we obtain
\begin{align}
    \label{eq:ap_noneq_Heff}
    &\left|\mathrm{Tr}_{F,S}\left[\Delta H_{S}(s_{1})\ket{\Psi(s_1;\theta )} \bra{\Psi(s_2;\theta )}\Delta H_{\mathrm{eff}}(s_{2})^{\dagger}\mathcal{V}\left(s_{1},s_{2}\right)^{\dagger}\right]\right|=\left|\Braket{\Psi(s_2;\theta )|\Delta H_{\mathrm{eff}}(s_{2})^{\dagger}\mathcal{V}\left(s_{1},s_{2}\right)^{\dagger}\Delta H_{S}(s_{1})|\Psi(s_1;\theta )}\right|\nonumber\\
    &\le\sqrt{\Braket{\Psi(s_1;\theta)|\Delta H_{S}(s_{1})^2|\Psi(s_1;\theta)}}\sqrt{\Braket{\Psi(s_2;\theta)|\Delta H_{\mathrm{eff}}(s_{2})^{\dagger}\Delta H_{\mathrm{eff}}(s_{2})|\Psi(s_2;\theta )}}=\sigma_{H_S}(s_{1})\sigma_{H_{\mathrm{eff}}}(s_{2}).
\end{align}
Here, we use Eq.~\eqref{eq:ap_VV}  (unitarity of $\mathcal{V}$) in the inequality  and use Eq.~\eqref{eq:ap_rho_VV} in the final equality. By combining this inequality with Eq.~\eqref{eq:ap2_def_mean}, Eq.~\eqref{eq:ap2_corr} and Eq.~\eqref{eq:ap2_second}, we obtain
\begin{align}
    \mathcal{K}_{s_{1}\neq s_{2}}\le \frac{2}{\theta^2}\int_{0}^{t}ds_{1}\sigma_{H_S}(s_{1})\int_{0}^{s_{1}}ds_{2} \sigma_{H_{\mathrm{eff}}}(s_{2})+\frac{1}{\theta^2}\left(\int_{0}^{t}ds \mathrm{Tr}_{S}\left[ H_{S}\rho_S(s)\right]\right)^2.
\end{align}
By combining Eq.~\eqref{eq:ap_tr_psi}, Eq.~\eqref{eq:ap_K_eq_result} and Eq.~\eqref{eq:ap_psi_dif_result} with this inequality, we obtain
\begin{align}
    \mathcal{B}(t)\le \overline{\mathcal{B}}(t) \equiv \mathcal{A}(t) + 8\int_{0}^{t}ds_{1}\sigma_{H_S}(s_{1})\int_{0}^{s_{1}}ds_{2} \sigma_{H_{\mathrm{eff}}}(s_{2}).
    \label{eq:ap_ub_result1}
\end{align}
From Eq.~\eqref{eq:ap2_def_stdev}, we obtain
\begin{align}
    \sigma_{H_{\mathrm{eff}}}(s)^2=\left\langle H_{S}^2+ \frac{1}{4}\left(\sum_m L_m^{\dagger}L_m\right)^2 +\frac{i}{2}\left[\sum_m L_m^{\dagger}L_m, H_{S}\right]\right\rangle(s) - \left(\langle H_{S}\rangle(s)\right)^2-\frac{1}{4}\left(\left\langle \sum_m L_m^{\dagger}L_m\right\rangle(s)\right)^2.
\end{align}
Applying the Robertson inequality $1/2\left|\langle[F,G]\rangle\right|\le \sigma_F\sigma_G$, we obtain
\begin{align}
    \sigma_{H_{\mathrm{eff}}}(s)^2\le \left|\frac{1}{2}\left\langle\left[\sum_m L_m^{\dagger}L_m, H_{S}\right]\right\rangle(s)\right| +\sigma_{H_S}^2(s)+\frac{1}{4}\sigma_{L^{\dagger}L}^2(s)\le \left(\sigma_{H_S}(s)+\frac{1}{2}\sigma_{L^{\dagger}L}(s)\right)^2,
\end{align}
where $\sigma_{L^{\dagger}L}(s)$ denotes the standard deviation of the operator $\sum_m L_m^{\dagger}L_m$.
Substituting this inequality into Eq.~\eqref{eq:ap_ub_result1}, we obtain
\begin{align}
    \overline{\mathcal{B}}(t)& \le \mathcal{A}(t)+4\left(\int_{0}^{t}ds\sigma_{H_S}(s)\right)^2+4\int_{0}^{t}ds_{1}\sigma_{H_S}(s_{1})\int_{0}^{s_{1}}ds_{2} \sigma_{L^{\dagger}L}(s_{2}).
    \label{eq:ap_ub_result2}
\end{align}
When the equality holds between the exact solution Eq.~\eqref{eq:main_result1} and the upper bound Eq.~\eqref{eq:BUB_def}, the first term of the right-hand side in Eq.~\eqref{eq:ap2_tr_expansion} must be non-negative from Eq.~\eqref{eq:ap2_corr}.
Furthermore, by the equality condition of the Cauchy-Schwarz inequality, it is necessary and sufficient for one of the following conditions to hold for $0\le s_{2} \le s_{1}\le t$ and $a(s_{1}, s_{2})\in[0,\infty)$. 
    \begin{enumerate}
        \item 
        \begin{align}
            \Delta H_S(s_{1}) \ket{\Psi(s_1;\theta)}&=a(s_{1}, s_{2})\mathcal{V}\left(s_{1},s_{2}\right)\Delta H_\mathrm{eff}(s_{2}) \ket{\Psi(s_2;\theta)},
            \label{eq:ap_eq_cond1}
        \end{align}
        \item 
        \begin{align}
        \mathcal{V}\left(s_{1},s_{2}\right)\Delta H_\mathrm{eff}(s_{2}) \ket{\Psi(s_2;\theta)}=0.
        \end{align}
    \end{enumerate}
\section{Specific case of exact representation\label{sec:specific_case}}
By letting $G_m\equiv [H_S, L_m]$ for $1\le m \le N_{C}$ and $r\in\mathbb{R}$, we assume that
\begin{align}
    \label{eq:commutation}
    G_m^{\dagger}L_m + L_m^{\dagger}G_m=2rH_{S}.
\end{align}
This condition agrees with the closed quantum system and the classical limit.
In this case, the exact solution of Eq.~\eqref{eq:main_result1} can be simplified as follows:
\begin{align}
    \mathcal{B}(t)=\mathcal{A}(t)+\frac{8}{r}\mathrm{Re}\left(\int_{0}^{t}ds\left(\exp\left(r(t-s)\right)-1\right)\mathrm{Tr}_S\left[H_{\mathrm{eff}}^{\dagger}H_{S}\rho_S\left(s\right)\right]\right)-4\left(\int_{0}^{t}ds\mathrm{Tr}_{S}\left[H_{S}\rho_{S}(s)\right]\right)^{2}.
    \label{eq:main_result1_simple}
\end{align}
Since $[H_{S}, L_m^{\dagger}]=-G_m^{\dagger}$, we have
\begin{align}
    \label{eq:H_LL}
    H_{S}L_m^{\dagger}L_m=L_m^{\dagger}H_{S}L_m -G_m^{\dagger}L_m=L_m^{\dagger}L_m H_{S}
    +L_m^{\dagger}G_m-G_m^{\dagger}L_m.
\end{align}
Using this relation for the Kraus operators
[Eqs.~\eqref{eq:Kraus_V0_def} and \eqref{eq:Kraus_Vm_def}]
we have
\begin{align}
    \label{eq:H_S_Kraus}
    \sum_m V_{m}(ds)^{\dagger} H_{S}V_{m}(ds)=H_{S}-\frac{1}{2}\sum_m\left(L_m^{\dagger}G_m-G_m^{\dagger}L_m\right)ds + \sum_m L_m^{\dagger}G_m ds+O(ds^2)=(1+rds)H_{S},
\end{align}
where we use Eq.~\eqref{eq:commutation}.
Generalizing Eq.~\eqref{eq:H_S_Kraus}, we obtain
\begin{align}
    \check{H}_{S}(s)=\exp(rs)H_{S}.
\end{align}
Substituting this relation into Eq.~\eqref{eq:main_result1}, the second term yields
\begin{align}
    &8\int_{0}^{t}ds_{1}\int_{0}^{s_{1}}ds_{2}\mathrm{Re}\left(\exp\left(r(s_{1}-s_{2})\right)\mathrm{Tr}_S\left[H_{\mathrm{eff}}^{\dagger}H_{S}\rho_S\left(s_{2}\right)\right]\right)=8\mathrm{Re}\left(\int_{0}^{t}ds_{2}\int_{s_{2}}^{t}ds_{1}\exp\left(r(s_{1}-s_{2})\right)\mathrm{Tr}_S\left[H_{\mathrm{eff}}^{\dagger}H_{S}\rho_S\left(s_{2}\right)\right]\right)\nonumber\\
    &=\frac{8}{r}\mathrm{Re}\left(\int_{0}^{t}ds\left(\exp\left(r(t-s)\right)-1\right)\mathrm{Tr}_S\left[H_{\mathrm{eff}}^{\dagger}H_{S}\rho_S\left(s\right)\right]\right).
\end{align}

\section{Derivation of Eq.~\eqref{eq:main_result1} from Ref.~\cite{Nakajima:2023:SLD}\label{sec:from_Nakajima}}
As mentioned in the main text, our main result [Eq.~\eqref{eq:main_result1}] can be derived from Ref.~\cite{Nakajima:2023:SLD},
which is elaborated in this section. 
Reference~\cite{Nakajima:2023:SLD} derived the following representation:
\begin{align}
    \label{eq:QDA_Nakajima}
    \mathcal{B}(t)&=\mathcal{A}(t)+ 4(I_1 + I_2)-4\left(\int_{0}^{t}ds\mathrm{Tr}_{S}\left[H_{S}\rho_{S}(s)\right]\right)^{2}, \\
     \label{eq:QDA_I1} 
    I_1&\equiv\int_{0}^{t}ds_{1}\int_{0}^{s_{1}}ds_{2} \mathrm{Tr}_S\left[\mathcal{K}_{2}\exp\left(\hat{\mathcal{L}}(s_{1}-s_{2})\right)\mathcal{K}_{1}\rho_S(s_{2})\right],\\
    \label{eq:QDA_I2}
    I_2&\equiv\int_{0}^{t}ds_{1}\int_{0}^{s_{1}}ds_{2} \mathrm{Tr}_S\left[\mathcal{K}_{1}\exp\left(\hat{\mathcal{L}}(s_{1}-s_{2})\right)\mathcal{K}_{2}\rho_S(s_{2})\right],
\end{align}
where $\hat{\bullet}$ denotes the vectorization of an operator~\cite{Landi:2023:CurFlucReview}, and $\mathcal{K}_1$ and $\mathcal{K}_2$ are the following superoperators:
\begin{align}
    \label{eq:def_K1}
    \mathcal{K}_{1}\bullet&=-iH_{\mathrm{eff}} \bullet +\frac12 \sum_k L_k \bullet L_k^{\dagger}, \\
    \label{eq:def_K2}
    \mathcal{K}_{2}\bullet&=i \bullet H_{\mathrm{eff}}^{\dagger}+\frac12 \sum_k L_k \bullet L_k^{\dagger}=(\mathcal{L}-\mathcal{K}_{1})\bullet.
\end{align}
For Eq.~\eqref{eq:def_K1} and Eq.~\eqref{eq:def_K2}, by using the cyclic property of the trace, we have
\begin{align}
    \label{eq:K1_another_rep}
    \mathrm{Tr}_{S}\left[\mathcal{K}_{1}\bullet\right]&=-i\mathrm{Tr}_{S}\left[H_S\bullet\right], \\
     \label{eq:K2_another_rep}
     \mathrm{Tr}_{S}\left[\mathcal{K}_{2}\bullet\right]&=i\mathrm{Tr}_{S}\left[H_S\bullet\right].
\end{align}
By representing $\exp\left(\hat{\mathcal{L}}(s-u)\right)$ using the Kraus operators 
[Eqs.~\eqref{eq:Kraus_V0_def} and \eqref{eq:Kraus_Vm_def}]
 and applying the cyclic property of the trace, we obtain
\begin{align}
    I_1&=i\int_{0}^{t}ds_{1}\int_{0}^{s_{1}}ds_{2} \mathrm{Tr}_S\left[\check{H}_S (s_{1}-s_{2})\mathcal{K}_{1}\rho_S(s_{2})\right]\nonumber\\
    &=\int_{0}^{t}ds_{1}\int_{0}^{s_{1}}ds_{2} \mathrm{Tr}_S\left[\check{H}_S (s_{1}-s_{2})H_{\mathrm{eff}}\rho_S(s_{2})\right]+\frac i2\int_{0}^{t}ds_{1}\int_{0}^{s_{1}}ds_{2} \sum_k \mathrm{Tr}_S\left[\check{H}_S (s_{1}-s_{2})L_k\rho_S(s_{2})L_k^{\dagger}\right].
\end{align}
where we use Eq.~\eqref{eq:def_K1}. Similarly, we have
\begin{align}
    I_2=\int_{0}^{t}ds_{1}\int_{0}^{s_{1}}ds_{2} \mathrm{Tr}_S\left[\check{H}_S (s_{1}-s_{2})\rho_S(s_{2})H_{\mathrm{eff}}^{\dagger}\right]-\frac i2\int_{0}^{t}ds_{1}\int_{0}^{s_{1}}ds_{2} \sum_k \mathrm{Tr}_S\left[\check{H}_S (s_{1}-s_{2})L_k\rho_S(s_{2})L_k^{\dagger}\right].
\end{align}
Since $\check{H}_S(s)$ is Hermite from Eq.~\eqref{eq:ap2_Hhat}, it follows that $\mathrm{Tr}_S\left[\check{H}_S (s_{1}-s_{2})H_{\mathrm{eff}}\rho_S(s_{2})\right]^*=\mathrm{Tr}_S\left[\check{H}_S (s_{1}-s_{2})\rho_S(s_{2})H_{\mathrm{eff}}^{\dagger}\right]$.
Hence, we have
\begin{align}
    I_1+I_2=2\int_{0}^{t}ds_{1}\int_{0}^{s_{1}}ds_{2}\mathrm{Re}\left(\mathrm{Tr}_S\left[H_{\mathrm{eff}}^{\dagger}\check{H}_{S}\left(s_{1}-s_{2}\right)\rho_S\left(s_{2}\right)\right]\right).
\end{align}

\section{Vectorization of exact solution\label{sec:vectorilization}}
In this section, we provide another representation of the quantum dynamical activity $\mathcal{B}(\tau)$ with a single integral using vectorization.

Let $\rho$ be the density operator represented as follows:
\begin{align}
    \rho = \sum_{i,j}\rho_{ij}\ket{i}\bra{j}.
    \label{eq:rho_def}
\end{align}
Using Choi-Jamio{\l}kowski isomorphism, 
$\rho$ can be represented as
\begin{align}
    \dblket{\rho}\equiv\sum_{i,j}\rho_{ij}\ket{j}\otimes\ket{i}.
    \label{eq:Liouville_rho_def}
\end{align}
Thus, it naturally follows that $\dblbra{\rho} \equiv \dblket{\rho}^\dagger$.
By adopting this notation, we can redefine the inner product to represent the Hilbert-Schmidt inner product as follows:
\begin{align}
    \mathrm{Tr}\left[A^{\dagger}B\right]=\dblbraket{A\mid B}.
\end{align}
Moreover, the following identity holds:
\begin{align}
    \dblket{ABC}=(C^{\top}\otimes A)\dblket{B}.
    \label{eq:ABC_dblket}
\end{align}
We assume that the matrix representation of the Lindblad superoperator is diagonalizable and assume that the steady state $\rho_{S}^{\mathrm{ss}}$ is unique. 
We define the left and right eigenvector as
\begin{align}
    \hat{\mathcal{L}}\dblket{x_j}=\lambda_j \dblket{x_j}, \\
    \dblbra{y_j}\hat{\mathcal{L}}=\dblbra{y_j}\lambda_j. 
\end{align}
These satisfy the following conditions:
\begin{align}
    \dblbraket{y_i|x_j}=\delta_{ij}.
\end{align}
For $\lambda_0=0$, we have $\dblbra{y_0}=\dblbra{\mathbb{I}}$ and $\dblket{x_0}=\dblket{\rho_{S}^{\mathrm{ss}}}$. Here, $\mathbb{I}$ is the identity operator in the vectorized space. 
The vectorized Lindblad superoperator is decomposed as follows:
\begin{align}
    \hat{\mathcal{L}}&=\sum_{j\neq 0} \lambda_j \dblket{x_j}\dblbra{y_j}.
\end{align}
The matrix exponential can be written as
\begin{align}
    \exp\left(\hat{\mathcal{L}}t\right) &= \dblket{\rho_{S}^{\mathrm{ss}}}\dblbra{\mathbb{I}} + \sum_{j\neq 0} \exp(\lambda_j t) \dblket{x_j}\dblbra{y_j}.
    \label{eq:decomp_exp}
\end{align}
We represent the second term of the right-hand side of  Eq.~\eqref{eq:main_result1}  using vectorization.
Using the Hermitian property of $\check{H}_{S}(s)$ and the cyclic property of the trace, we can write the second term as
\begin{align}
    8\int_{0}^{t}ds_{1}\int_{0}^{s_{1}}ds_{2}\mathrm{Re}\left(\mathrm{Tr}_S\left[\check{H}_{S}\left(s_{1}-s_{2}\right)H_{\mathrm{eff}}\rho_S\left(s_{2}\right)\right]\right).
\end{align}
Using the cyclic property of the trace and applying the Kraus operators to $H_{\mathrm{eff}}\rho_S(s_{2})$, we obtain 
\begin{align}
    8\int_{0}^{t}ds_{1}\int_{0}^{s_{1}}ds_{2}\mathrm{Re}\dblbra{\mathbb{I}}\hat{H}_{S}\exp\left(\mathcal{\hat{L}}(s_{1}-s_{2})\right)\hat{H}_{\mathrm{eff}}\dblket{\rho_S(s_{2}}.
\end{align}
Substituting Eq.~\eqref{eq:decomp_exp} into this equation and changing the order of the integral, we have
\begin{align}
    &8\int_{0}^{t}ds_{2}\int_{s_{2}}^{t}ds_{1}\mathrm{Re}\dblbra{\mathbb{I}}\hat{H}_{S}\exp\left(\mathcal{\hat{L}}(s_{1}-s_{2})\right)\hat{H}_{\mathrm{eff}}\dblket{\rho_S(s_{2}}\nonumber \\
    &=8\mathrm{Tr}_S[H_{S}\rho_{S}^{\mathrm{ss}}]\int_{0}^{t}ds(t-s) \mathrm{Tr}_S[H_{S}\rho_S(s)]+8\int_{0}^{t}ds_{2}\int_{s_{2}}^{t}ds_{1}\mathrm{Re}\dblbra{\mathbb{I}}\hat{H}_{S}\sum_{j\neq 0} \exp\left(\lambda_j(s_{1}-s_{2})\right) \dblket{x_j}\dblbra{y_j}\hat{H}_{\mathrm{eff}}\dblket{\rho_S(s_{2}}\nonumber\\
    &=8\mathrm{Tr}_S[H_{S}\rho_{S}^{\mathrm{ss}}]\int_{0}^{t}ds(t-s) \mathrm{Tr}_S[H_{S}\rho_S(s)]+8\mathrm{Re}\int_{0}^{t}ds\dblbra{\mathbb{I}}\hat{H}_{S}\sum_{j\neq 0} \frac{1}{\lambda_j}\left(\exp\left(\lambda_j\left(t-s\right)\right)-1\right)\dblket{x_j}\dblbra{y_j}\hat{H}_{\mathrm{eff}}\dblket{\rho_S(s)}\nonumber\\
    &=8\mathrm{Tr}_S[H_{S}\rho_{S}^{\mathrm{ss}}]\int_{0}^{t}ds(t-s) \mathrm{Tr}_S[H_{S}\rho_S(s)]+8\mathrm{Re}\int_{0}^{t}ds\dblbra{\mathbb{I}}\hat{H}_{S}\hat{\mathcal{L}}^\mathrm{D}\left(\exp\left(\hat{\mathcal{L}}(t-s)\right)-1\right)\hat{H}_{\mathrm{eff}}\dblket{\rho_S(s)},
    \label{eq:vec_result1}
\end{align}
where 
\begin{align}
    \hat{\mathcal{L}}^\mathrm{D} &\equiv \sum_{j\neq 0} \frac{1}{\lambda_j} \dblket{x_j}\dblbra{y_j}.
    \label{eq:Drazin_inverse_def}
\end{align}
$\hat{\mathcal{L}}^\mathrm{D}$ is the Drazin pseudo-inverse, which satisfies 
\begin{align}
    \label{eq:drazin_projection}\hat{\mathcal{L}}^\mathrm{D}\hat{\mathcal{L}}&=\hat{\mathcal{L}}\hat{\mathcal{L}}^\mathrm{D}=\mathcal{P},\\
    \label{eq:drazin_moore}
    \hat{\mathcal{L}}^\mathrm{D}&=\mathcal{P}\hat{\mathcal{L}}^{+}\mathcal{P},
\end{align}
where $\hat{\mathcal{L}}^{+}$ denotes the Moore-Penrose pseudo-inverse, and 
\begin{align}
    \mathcal{P}&\equiv \mathbb{I}-\dblket{\rho_{S}^{\mathrm{ss}}}\dblbra{\mathbb{I}}=\sum_{j\neq 0} \dblket{x_j}\dblbra{y_j}.
\end{align}
By combining  Eq.~\eqref{eq:main_result1}  and Eq.~\eqref{eq:vec_result1}, we obtain
\begin{align}
   \mathcal{B}(t)&=\mathcal{A}(t)+8\mathrm{Tr}_S[H_{S}\rho_{S}^{\mathrm{ss}}]\int_{0}^{t}ds(t-s) \mathrm{Tr}_S[H_{S}\rho_S(s)]+8\mathrm{Re}\int_{0}^{t}ds\dblbra{\mathbb{I}}\hat{H}_{S}\hat{\mathcal{L}}^\mathrm{D}\left(\exp\left(\hat{\mathcal{L}}(t-s)\right)-1\right)\hat{H}_{\mathrm{eff}}\dblket{\rho_S(s)}\nonumber\\
   &-4\left(\int_{0}^{t}ds\mathrm{Tr}_{S}\left[H_{S}\rho_{S}(s)\right]\right)^{2}.
    \label{eq:sup_exact_QDA_vec}
\end{align}
\section{Approximations of exact solution\label{sec:approximation}}
We provide two approximations when $t$ is either small or large. 
Let $\nu \equiv \min_{j>0} |\mathrm{Re}\lambda_j|$ and let
    \begin{align}
        \gamma(t)\equiv \max\left(\frac{1}{t}\sum_{j\neq 0} \frac{1}{|\lambda_j|},\; \exp(-\nu t)\right).
    \end{align}  
We first provide an approximation of the quantum dynamical activity  when $\gamma(t)\ll 1$.  
From $\mathbb{I}=\dblket{\rho_{S}^{\mathrm{ss}}}\dblbra{\mathbb{I}} + \sum_{j\neq 0}  \dblket{x_j}\dblbra{y_j}$, we have
\begin{align}
    \dblket{\rho_S(t)}=\dblket{\rho_{S}^{\mathrm{ss}}} + \sum_{j\neq 0} \exp(t\lambda_j) \dblket{x_j}\dblbraket{y_j|\rho_S(0)}.
    \label{eq:rho_expansion}
\end{align}
From Eq.~\eqref{eq:rho_expansion}, the second term in Eq.~\eqref{eq:vec_result1} can be written as 
\begin{align}
    &-8t\mathrm{Re}\dblbra{\mathbb{I}}\hat{H}_{S}\hat{\mathcal{L}}^\mathrm{D}\hat{H}_{\mathrm{eff}}\dblket{\rho_{S}^{\mathrm{ss}}}+8\mathrm{Re}\dblbra{\mathbb{I}}\hat{H}_{S}\left(\hat{\mathcal{L}}^\mathrm{D}\right)^2\left(1-\exp\left(\hat{\mathcal{L}}t\right)\right)\hat{H}_{\mathrm{eff}}\dblket{\rho_\mathrm{ss}}\nonumber \\
    &+8\mathrm{Re}\int_{0}^{t}ds\sum_{j, k\neq 0} \frac{1}{\lambda_j}\left(\exp\left(\lambda_j t +(\lambda_k-\lambda_j) s\right)-\exp(\lambda_k s)\right)\dblbra{\mathbb{I}}\hat{H}_{S}\dblket{x_j}\dblbra{y_j}\hat{H}_{\mathrm{eff}}\dblket{x_k}\dblbraket{y_k|\rho_S(0)}.
    \label{eq:second_QDA}
\end{align}
By integrating the third term in Eq.~\eqref{eq:second_QDA}, we obtain
\begin{align}
    &8\mathrm{Re}\sum_{j\neq 0} \left(\frac{t\exp(\lambda_j t)}{\lambda_j}-\frac{1}{\lambda_j^2}(\exp(\lambda_J t)-1)\right)\dblbra{\mathbb{I}}\hat{H}_{S}\dblket{x_j}\dblbra{y_j}\hat{H}_{\mathrm{eff}}\dblket{x_j}\dblbraket{y_j|\rho_S(0)}\nonumber \\
    &+8\mathrm{Re}\sum_{j\neq k;\; j,k\neq 0} \left(\frac{1}{\lambda_j (\lambda_k-\lambda_j)}\left(\exp(\lambda_k t) -\exp(\lambda_j t)\right)-\frac{1}{\lambda_j \lambda_k}(\exp(\lambda_k t)-1)\right)\dblbra{\mathbb{I}}\hat{H}_{S}\dblket{x_j}\dblbra{y_j}\hat{H}_{\mathrm{eff}}\dblket{x_k}\dblbraket{y_k|\rho_S(0)}.
    \label{eq:second_QDA2}
\end{align}
By combining Eq.~\eqref{eq:second_QDA} and Eq.~\eqref{eq:second_QDA2}, the second term in Eq.~\eqref{eq:vec_result1} is expressed as
    \begin{align}       
        -8t\mathrm{Re}\dblbra{\mathbb{I}}\hat{H}_{S}\hat{\mathcal{L}}^\mathrm{D}\hat{H}_{\mathrm{eff}}\dblket{\rho_{S}^{\mathrm{ss}}} + tO(\gamma(t)).
    \end{align}
Hence, we have
    \begin{align}
        \mathcal{B}(t)= \mathcal{A}(t)+ t\mathcal{Z}+8\mathrm{Tr}_S[H_{S}\rho_{S}^{\mathrm{ss}}]\int_{0}^{t}ds(t-s) \mathrm{Tr}_S[H_{S}\rho_S(s)]-4\left(\int_{0}^{t}ds\mathrm{Tr}_{S}\left[H_{S}\rho_{S}(s)\right]\right)^{2}+O(t\gamma(t)),
        \label{eq:vec_result2}
    \end{align}
where 
\begin{align}
    \mathcal{Z}\equiv -8t\mathrm{Re}\dblbra{\mathbb{I}}\hat{H}_{S}\hat{\mathcal{L}}^\mathrm{D}\hat{H}_{\mathrm{eff}}\dblket{\rho_{S}^{\mathrm{ss}}}.
    \label{eq:def_Z}
\end{align}
Note that the degree of $8\mathrm{Tr}_S[H_{S}\rho_{S}^{\mathrm{ss}}]\int_{0}^{t}ds(t-s) \mathrm{Tr}_S[H_{S}\rho_S(s)]-4\left(\int_{0}^{t}ds\mathrm{Tr}_{S}\left[H_{S}\rho_{S}(s)\right]\right)^{2}$ is at most one with respect to $t$ from Eq.~\eqref{eq:rho_expansion}.
When $\rho_S(0)=\rho_{S}^{\mathrm{ss}}$, the equation can be simplified as 
    \begin{align}
        \mathcal{B}(t)= \mathcal{A}(t)+ t\mathcal{Z} + O(t\gamma(t)).
        \label{eq:vec_result2}
    \end{align}
Next, we provide an approximation when we can ignore $O(t^3)$.
From $\exp(\hat{\mathcal{L}}t)-1\sim \hat{\mathcal{L}}t$ and Eq.~\eqref{eq:drazin_projection}, the equation  Eq.~\eqref{eq:vec_result1} yields
\begin{align}
    &8\mathrm{Tr}_S[H_{S}\rho_{S}^{\mathrm{ss}}]\int_{0}^{t}ds(t-s) \mathrm{Tr}_S[H_{S}\rho_S(s)]+8\mathrm{Re}\int_{0}^{t}ds(t-s)\dblbra{\mathbb{I}}\hat{H}_{S}\mathcal{P}\hat{H}_{\mathrm{eff}}\dblket{\rho_S(s)}=8\mathrm{Re}\int_{0}^{t}ds(t-s)\dblbra{\mathbb{I}}\hat{H}_{S}\hat{H}_{\mathrm{eff}}\dblket{\rho_S(s)}\nonumber\\
    &=8\mathrm{Re}\int_{0}^{t}ds(t-s) \mathrm{Tr}_S\left[H_SH_{\mathrm{eff}}\rho_S(s)\right].
\end{align}
By substituting this equation into Eq.~\eqref{eq:sup_exact_QDA_vec}, we obtain 
\begin{align}
     \mathcal{B}(t)= \mathcal{A}(t)+8\mathrm{Re}\int_{0}^{t}ds(t-s) \mathrm{Tr}_S\left[H_SH_{\mathrm{eff}}\rho_S(s)\right]-4\left(\int_{0}^{t}ds\mathrm{Tr}_{S}\left[H_{S}\rho_{S}(s)\right]\right)^{2}+O(t^3).
\end{align}
Furthermore, from $\rho_S(t)=\rho_S(0)+O(t)$, we obtain
\begin{align}
     \mathcal{B}(t)= \mathcal{A}(t)+4t^2\mathrm{Re} \mathrm{Tr}_S\left[\Delta H_S \Delta H_{\mathrm{eff}}\rho_S(0)\right]+O(t^3).
\end{align}
Since the second term includes $4t^2\sigma_{H_S}^2(0)$, this equation corresponds to the upper bound Eq.~\eqref{eq:BUB_def}.

\section{Asymptotic expression of $\mathcal{B}(\tau)$\label{sec:asymp_approach}}

In Ref.~\cite{Hasegawa:2020:QTURPRL}, the quantum dynamical activity $\mathcal{B}(\tau)$ was evaluated within the limit of $\tau \to \infty$.
In this section, we review its representation. 

Suppose that the Lindblad equation has a steady-state solution. 
Using vectorization, the quantum dynamical activity for $\tau \to \infty$ can be represented as
\begin{align}
   \mathcal{B}_{\mathrm{\infty}}(\tau)\equiv\tau(\mathfrak{a}+\mathfrak{b}_{c})=\tau\left(\mathfrak{a}+4\mathcal{Z}_{1}+4\mathcal{Z}_{2}\right),
    \label{eq:Bt_long_tau_app}
\end{align}
where $\mathcal{Z}_1$ and $\mathcal{Z}_2$ are defined as
\begin{align}
    \mathcal{Z}_{1}&=-\dblbra{\mathbb{I}}\hat{\mathcal{K}}_{1}(I-\dblket{\rho_{S}^{\mathrm{ss}}}\dblbra{\mathbb{I}})\hat{\mathcal{L}}^{+}(I-\dblket{\rho_{S}^{\mathrm{ss}}}\dblbra{\mathbb{I}})\hat{\mathcal{K}}_{2}\dblket{\rho_{S}^{\mathrm{ss}}},\label{eq:mathcalZ1_def}\\\mathcal{Z}_{2}&=-\dblbra{\mathbb{I}}\hat{\mathcal{K}}_{2}(I-\dblket{\rho_{S}^{\mathrm{ss}}}\dblbra{\mathbb{I}})\hat{\mathcal{L}}^{+}(I-\dblket{\rho_{S}^{\mathrm{ss}}}\dblbra{\mathbb{I}})\hat{\mathcal{K}}_{1}\dblket{\rho_{S}^{\mathrm{ss}}}.
    \label{eq:mathcalZ2_def}
\end{align}
Here, $+$ denotes the  Moore-Penrose pseudoinverse, and
$I$ is the identity operator in the vectorized space. 
Using the Drazin inverse $\hat{\mathcal{L}}^\mathrm{D}$ given by Eq.~\eqref{eq:Drazin_inverse_def}, we obtain
\begin{align}
    \mathcal{Z}_{1}&=-\dblbra{\mathbb{I}}\hat{\mathcal{K}}_{1}\hat{\mathcal{L}}^{\mathrm{D}}\hat{\mathcal{K}}_{2}\dblket{\rho_{S}^{\mathrm{ss}}},\label{eq:mathcalZ1_drazin}\\\mathcal{Z}_{2}&=-\dblbra{\mathbb{I}}\hat{\mathcal{K}}_{2}\hat{\mathcal{L}}^{\mathrm{D}}\hat{\mathcal{K}}_{1}\dblket{\rho_{S}^{\mathrm{ss}}}.\label{eq:mathcalZ2_drazin}
\end{align}

\subsection{Equivalence of Eq.~\eqref{eq:vec_result2} and Eq.~\eqref{eq:Bt_long_tau_app}}
We show that Eq.~\eqref{eq:vec_result2} and Eq.~\eqref{eq:Bt_long_tau_app} are equivalent.
From Eq.~\eqref{eq:def_K1}, Eq.~\eqref{eq:def_K2}, and $\mathcal{L}\rho_{S}^{\mathrm{ss}}=0$, we have
\begin{align}
    \mathcal{K}_{1}\rho_{S}^{\mathrm{ss}}&=-\frac i2 H_{\mathrm{eff}}\rho_{S}^{\mathrm{ss}}-\frac i2\rho_{S}^{\mathrm{ss}}H_{\mathrm{eff}}^{\dagger},\label{eq:mathcalK1_def}
    \\\mathcal{K}_{2}\rho_{S}^{\mathrm{ss}}&
    =-\mathcal{K}_{1}\rho_{S}^{\mathrm{ss}}.
    \label{eq:mathcalK2_def}
\end{align}
From Eq.~\eqref{eq:K1_another_rep}, we obtain
\begin{align}
    \dblbra{\mathbb{I}}\hat{\mathcal{K}}_{1}\bullet\dblket{\rho_{S}^{\mathrm{ss}}}&=\mathrm{Tr}_S[\mathcal{K}_{1}\bullet \rho_{S}^{\mathrm{ss}}]=-i\dblbra{\mathbb{I}} \hat{H}_S \bullet\dblket{\rho_{S}^{\mathrm{ss}}}.
\end{align}
By combining these relations with Eq.~\eqref{eq:mathcalZ1_drazin}, we obtain 
\begin{align}
    \mathcal{Z}_{1}&=-\frac{1}{2}\dblbra{\mathbb{I}}\hat{H}_S \hat{\mathcal{L}}^\mathrm{D}\hat{H}_{\mathrm{eff}}\dblket{\rho_{S}^{\mathrm{ss}}}-\frac{1}{2}\dblbra{\mathbb{I}}\hat{H}_S \hat{\mathcal{L}}^\mathrm{D}\dblket{\rho_{S}^{\mathrm{ss}}H_{\mathrm{eff}}^\dagger}.
\end{align}
From Eq.~\eqref{eq:K2_another_rep} and Eq.~\eqref{eq:mathcalK2_def}, we obtain
\begin{align}
    \mathcal{Z}_{2}=\mathcal{Z}_{1}.
\end{align}
From $\hat{\mathcal{L}}^\mathrm{D}=\int_0^\infty dt \exp\left(\hat{\mathcal{L}} t\right)$
and the cyclic property of the trace, and by acting $\exp\left(\hat{\mathcal{L}} t\right)$ to $H_S$, we obtain
\begin{align}
    \left(\dblbra{\mathbb{I}}\hat{H}_{S}\hat{\mathcal{L}}^\mathrm{D}\hat{H}_{\mathrm{eff}}\dblket{\rho_{S}^{\mathrm{ss}}}\right)^*=\int_0^\infty dt \mathrm{Tr}_S[\check{H}_S(t)H_{\mathrm{eff}}\rho_{S}^{\mathrm{ss}}]^*=\int_0^\infty dt \mathrm{Tr}_S[\check{H}_S(t)\rho_{S}^{\mathrm{ss}}H_{\mathrm{eff}}^\dagger]=\dblbra{\mathbb{I}}\hat{H}_{S}\hat{\mathcal{L}}^\mathrm{D}\dblket{\rho_{S}^{\mathrm{ss}}H_{\mathrm{eff}}^\dagger},
\end{align}
where we use the Hermitian of $\check{H}_S(t)$.
Therefore, we obtain $\mathcal{Z}=4\mathcal{Z}_{1}+4\mathcal{Z}_{2}$.
    
\end{widetext}

\end{document}